\documentclass[usenatbib]{mn2e} 
\usepackage[utf8]{inputenc}
\usepackage{natbib,epsfig,cite,mystyle,amsmath,amssymb,graphicx}
\usepackage{array, xcolor, caption}
\usepackage{hyperref}
\hypersetup{colorlinks,allcolors=black}

\usepackage{bm}
\usepackage{graphicx,color}
\usepackage{latexsym,amsmath,amssymb,graphicx,booktabs}
\usepackage{placeins}
\usepackage{subfigure}
\usepackage{amsmath}
\usepackage{mathtools}
\usepackage{appendix}
\usepackage[normalem]{ulem}

\definecolor{MyBlue}{rgb}{0.15,0.15,0.70}
\definecolor{Dgreen}{rgb}{0,0.7,0.0}

\hypersetup{
colorlinks=true,
citecolor=MyBlue,
linkcolor=MyBlue,
urlcolor=MyBlue
}

\usepackage{amssymb}
\usepackage{amsmath}
\usepackage{amsfonts}
\usepackage{upgreek}
\usepackage{latexsym}
\usepackage{appendix}
\usepackage{dsfont}
\usepackage{orcidlink}

\newcommand\ees{\end{eqnarray}}
\newcommand\bees{\begin{eqnarray}}

\usepackage[export]{adjustbox}

\newcommand\spart{\;\raise1.0pt\hbox{/}\hskip-6pt\partial}
\newcommand\spartb{\;\overline{\raise1.0pt\hbox{/}\hskip-6pt\partial}}

\newcommand{\be}{\begin{equation}}
\newcommand{\ee}{\end{equation}}
\newcommand{\bk}{{\bm k}}

\newcommand{\beqa}{\begin{eqnarray}}
\newcommand{\eeqa}{\end{eqnarray}}

\newcommand{\bw}{{\mathbf{w}}}
\newcommand{\bv}{{\mathbf{v}}}

\newcommand{\nn}{\nonumber}

\newcommand{\tR}{\tilde{R}}
\newcommand{\tA}{\tilde{A}}
\newcommand{\thh}{\tilde{h}}
\newcommand{\te}{\tilde{e}}
\newcommand{\tn}{\tilde{n}}
\newcommand{\ttheta}{\tilde{\theta}}
\newcommand{\tphi}{\tilde{\phi}}
\newcommand{\tbA}{\tilde{\mathbf{A}}}
\newcommand{\tbn}{\tilde{\mathbf{n}}}
\newcommand{\bn}{\mathbf{n}}

\newcommand{\bbA}{\mathbf{A}}
\begin{document}
\title[ ]{Aberration of gravitational waveforms by peculiar velocity}
\author[Bonvin et al.]{Camille~Bonvin$^{1}$\thanks{camille.bonvin@unige.ch}\orcidlink{0000-0002-5318-4064},
Giulia~Cusin$^{1,2}$
\orcidlink{0000-0001-6046-1237},
Cyril~Pitrou$^{2}$\orcidlink{0000-0002-1747-7847}, 
Simone~Mastrogiovanni$^{3,4}$\orcidlink{0000-0003-1606-4183},
\newauthor 
Giuseppe~Congedo$^5$\orcidlink{000-0003-2508-0046}
and Jonathan~Gair$^6$\orcidlink{0000-0002-1671-3668}
\\
$^{1}$Universit\'e de Gen\`eve, D\'epartement de Physique Th\'eorique and Centre for Astroparticle Physics, 24 quai Ernest-Ansermet, \\CH-1211 Gen\`eve 4, Switzerland\\
$^{2}$Sorbonne Université, CNRS, UMR 7095, Institut d'Astrophysique de Paris, 75014 Paris, France \\
$^{3}$ Artemis, Université Côte d’Azur, Observatoire de la Côte d’Azur, CNRS, F-06304 Nice, France\\
$^{4}$ INFN, Sezione di Roma, I-00185 Roma, Italy\\
$^{5}$ Institute for Astronomy, School of Physics and Astronomy, University of Edinburgh, Royal Observatory, \\
Blackford Hill, Edinburgh, EH9 3HJ, United Kingdom, \\
$^6$ Max-Planck-Institut f\"ur Gravitationsphysik, Albert-Einstein-Institut,  Am M\"uhlenberg 1, 14476 Potsdam-Golm, Germany 
}

\pagerange{\pageref{firstpage}--\pageref{lastpage}} \pubyear{2022}
\maketitle
\label{firstpage}

\begin{abstract}
One key prediction of General Relativity is that gravitational waves are emitted with a pure spin-2 polarisation. Any extra polarisation mode, spin-1 or spin-0, is consequently considered a smoking gun for deviations from General Relativity. In this paper, we show that the velocity of merging binaries with respect to the observer gives rise to spin-1 polarisation in the observer frame even in the context of General Relativity. These are pure projection effects, proportional to the plus and cross polarisations in the source frame, hence they do not correspond to new degrees of freedom. We demonstrate that the spin-1 modes can always be rewritten as pure spin-2 modes coming from an aberrated direction. Since gravitational waves are not isotropically emitted around binary systems, this aberration modifies the apparent orientation of the binary system with respect to the observer: the system appears slightly rotated due to the source velocity. 
Fortunately, this bias does not propagate to other parameters of the system (and therefore does not spoil tests of General Relativity), since the impact of the velocity can be fully reabsorbed into new orientation angles. 
\end{abstract}


\section{Introduction}

Binary systems of compact objects, like neutron stars or black holes, are predicted by General Relativity to emit gravitational waves (GW) with spin-2 polarisations. These spin-2 modes have been observed for the first time by the interferometer LIGO and Virgo in 2015~\citep{LIGOScientific:2016aoc} and from subsequent GW events \citep{2019PhRvX...9c1040A, 2021PhRvX..11b1053A,2021arXiv210801045T,2021arXiv211103606T}. From a theoretical point of view it is of crucial importance to model the expected signal as precisely as possible, in order to use these GW events to probe, on one hand, the physics of binary systems \citep{2021hgwa.bookE..39C}, and, on the other hand, the validity of General Relativity \citep{2019PhRvD.100j4036A,2021PhRvD.103l2002A,2021arXiv211206861T}. A lot of effort has been devoted to calculate GW waveforms accounting for the relative velocity of the two objects in the binary, up to high order in the post-Newtonian expansion, see e.g.~\citet{Blanchet:2013haa,2021hgwa.bookE..31I,2021hgwa.bookE..32S,2021hgwa.bookE..34Z} for a more recent review. However, these frameworks usually neglect the fact that the centre of mass of the binary is itself moving with respect to the observer, due to the gravitational interaction with the host galaxy, host cluster, and the large-scale structure of the Universe. 

Recently several studies have started exploring the effect of the binary peculiar velocity on the waveform of a GW signal. In particular, it has been found that the variation of the velocity during the time of observation modifies the waveform in a non-negligible way, an effect that is relevant for an interferometer like LISA, that will follow GW signals during months and even years, see e.g. \citet{Bonvin:2016qxr, Tamanini_2020, Toubiana:2020drf, Sberna:2022qbn}. Other authors have addressed the impact of the binary peculiar motion for cosmological studies \citep{2021A&A...646A..65M}.
However, all these works focus on kinematic distortions of the amplitude and phase of the wave, assuming that the two emitted polarisations are affected in the same way by kinematic effects, i.e.\ effectively neglecting the spin-2 (tensorial) nature of the wave and treating the two wave polarisations as scalar waves. This is of course an approximation, as we know that a GW is in fact a spin-2 quantity, which consequently transforms as a rank-2 tensor under a Lorentz boost. 

\begin{figure}
    \centering
    \includegraphics[scale=0.26]{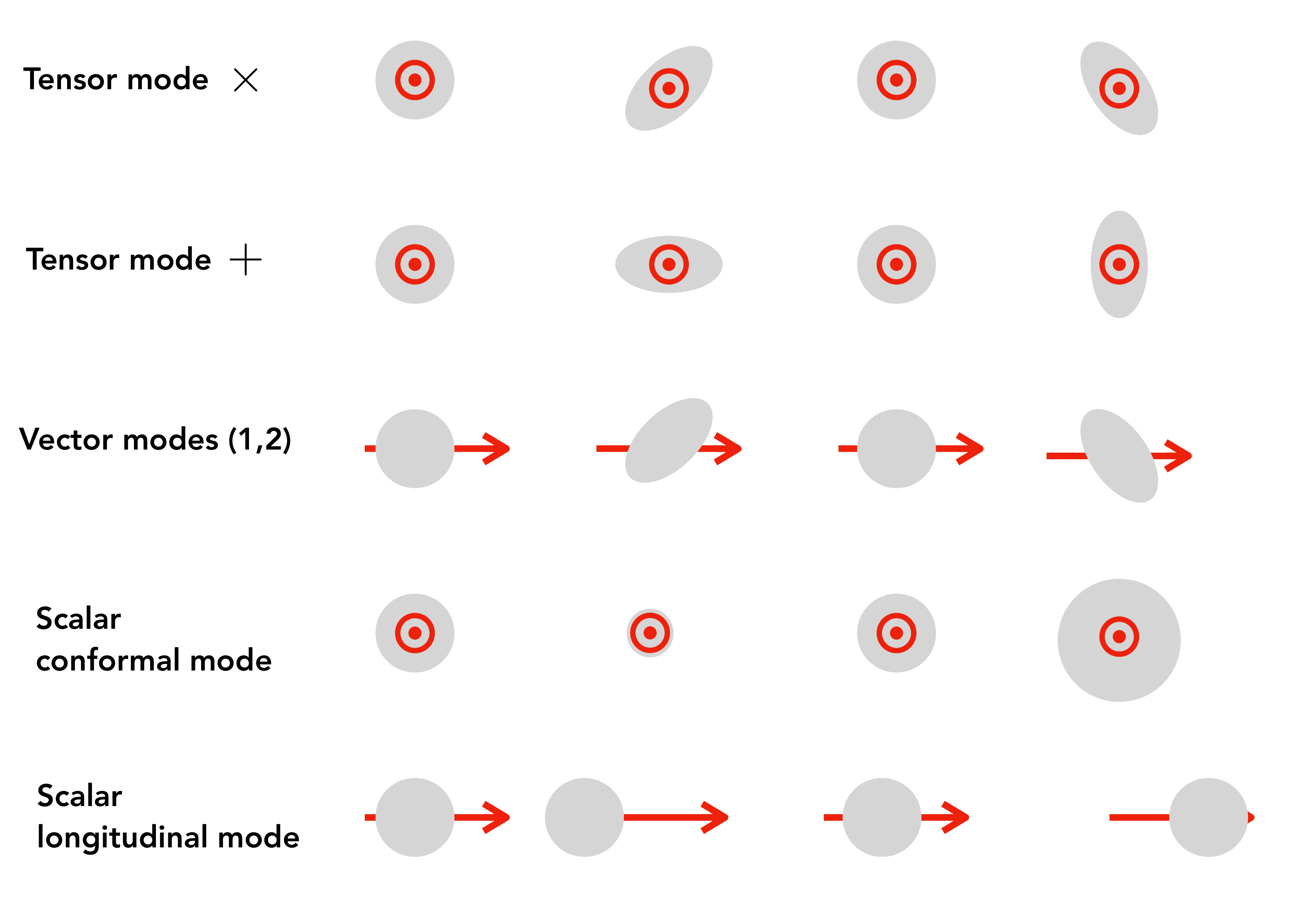}
    \caption{\label{Polarisations} Effect of different wave polarisations on a sphere of test-particles. The arrows indicate the direction of propagation of the wave. Adapted from \citet{deRham:2014zqa}.}
\end{figure}

In this paper, we study the effect of the binary peculiar velocity on the observed signal, accounting for the full polarisation structure of the GW. 
We show that the component of the binary velocity orthogonal to the line of sight (hereafter \emph{transverse} velocity) changes the antenna pattern of an interferometer, generating spin-1 modes. Moreover, for a network of detectors, the transverse velocity changes also the time delay between interferometers (or similarly the phase shift). In Fig.\,\ref{Polarisations} we plot the effect of various wave polarisations on a ring of test particles. Spin-1 polarisations give a vectorial deformation of the ring along the direction of propagation of the wave. 

We then show that since the spin-1 modes have the same time dependence as the spin-2 modes, they can always be rewritten as spin-2 modes coming from an aberrated direction, and with a mixing of the two polarisations. Moreover, we show that the time delay between different interferometers can be rewritten in terms of the same aberrated direction. This means that only the aberrated direction and the aberrated polarisations can be measured. Importantly, the aberrated direction which allows us to re-absorb spin-1 modes into spin-2 modes is the same as the aberrated direction inferred from the propagation of electromagnetic signals emitted by moving sources. As a consequence, detecting a luminous counterpart would not help in reconstructing the spin-1 modes, nor measuring the binary transverse velocity.

Since GW emission is not isotropic, aberration and the mixing of polarisations have a direct impact on the amplitude of the detected signal. When reconstructing the parameters of the binary system from the detected signal, we find that the angles describing the orientation of the binary system are biased by the transverse peculiar velocity: the system appears rotated with respect to the observer. Fortunately, this effect has no impact on the other parameters of the system, like the luminosity distance or the chirp mass,\footnote{These quantities are of course affected by the longitudinal component of the velocity through Doppler effects, but there are not affected by the transverse velocity which aberrates the signal and mixes the polarisations.} since the transverse velocity can be fully reabsorbed into an aberrated direction and mixed polarisations. As a consequence transverse velocities do not invalidate reconstruction of cosmological and astrophysical parameters with GWs.

The rest of the paper is structured as follows: after an overview of general concepts in Section~\ref{sec:general}, we present a detailed derivation of velocity-induced effects on the polarisation structure of the wave in Section \ref{theory}. In Section~\ref{sec:aberrated} we show how spin-1 components can be re-written as spin-2 components coming from an aberrated direction and in Section~\ref{sec:time delay} we demonstrate that the time-delay is proportional to this same aberrated direction. In Section~\ref{sec:obs} we discuss the observational impact of aberration and we conclude in Section~\ref{sec:conclusion}, where we discuss differences with \citet{Torres-Orjuela:2018ejx} and \citet{Torres-Orjuela:2020dhw}. Technical derivations are presented in a series of appendices.

{\bf{Notation:}} we work with units where the speed of light is set to one,  $c=1$. With $v_1$ and $v_2$ we denote projections of the source peculiar velocity orthogonal to the line of sight, and with $v_3$ the component along the line of sight. We denote with a tilde quantities in the source frame, and without tilde quantities in the observer frame (that we will also refer to as \emph{aberrated frame}). Moreover, since typically we expect peculiar velocities to be non-relativistic, in our computation we keep only linear order terms in $v/c$. This leads to simpler formulas whose physical interpretation is more transparent.

\section{General concepts}
\label{sec:general}

We start by reviewing the standard result of detector armlength variation induced by an incoming GW emitted by a source which is at rest with respect to the interferometer. This will serve as a basis for Section~\ref{theory} where we show how this derivation is modified if the source is moving. We consider two test particles in free fall, i.e.\ moving on nearby geodesics. The vector connecting these two geodesics, $\xi^\mu$, obeys the geodesic deviation equation 
\be
\frac{D^2\xi^{\mu}}{D\tau^2}=-\mathcal{R}^{\mu}_{\,\,\nu\rho\sigma}\xi^{\rho}u^{\nu}u^{\sigma}\,, \label{eq:deviation}
\ee
where $\tau$ is the proper time of the particles and $u^{\mu}$ is their four-velocity. In the frame of the particles, which we call ``observer frame", we have by construction $u^{\mu}=(-1, 0, 0, 0)$, and the geodesic deviation equation becomes 
\be\label{geo}
\frac{d^2\xi^i}{dt^2}=- S_{ij} \xi^j\,.
\ee
Here $t$ is the coordinate time, which is related to the proper time by $u^0=-dt/d\tau=-1$, and $S$ is the 
\emph{driving force matrix}, defined as 
\be\label{driving}
S_{ij}\equiv \mathcal{R}_{0i0j}\,,
\ee
where $i,j$ span the spatial coordinates $x,y,z$.

The passage of a GW affects the Riemann tensor and consequently the driving force matrix. We perturb the Minkowski metric as $g_{\mu\nu} = \eta_{\mu\nu} + h_{\mu\nu}$. At linear order in the perturbation $h_{\mu\nu}$, the Riemann tensor is given by $\mathcal{R}^{\mu\nu}_{\phantom{\mu\nu}\rho\sigma} =-2 \partial^{[\mu} \partial_{[\rho} h^{\nu]}_{\,\,\sigma]}$ which is manifestly invariant under an infinitesimal gauge transformation $g_{\mu\nu} \to g_{\mu\nu} - \partial_{\mu} \xi_{\nu} - \partial_{\nu} \xi_{\mu}$. From \eqref{driving} the linear order driving force matrix reads
\begin{align}\label{eq:driving}
S_{ij}=\frac{1}{2}\left(-\partial_{i}\partial_{j}h_{00}+
\partial_{0}\partial_{j}h_{0i}-\partial_{0}\partial_{0}h_{ij}+\partial_{i}\partial_{0}h_{0j}\right)\,.
\end{align}
Usually, one assumes that the source emitting a GW is at rest with respect to the observer. The metric in the wave zone  can then be written in the transverse traceless (TT) gauge and the driving force matrix reduces to
\be\label{Relectric}
S_{ij}=-\frac{1}{2}\ddot{h}_{ij}^{\rm TT}\,,
\ee 
leading to
\be\label{geo2}
\frac{d^2\xi^i}{dt^2}=\frac{1}{2}\ddot{h}_{ij}^{\rm TT} \xi^j\,,
\ee
with dots denoting differentiation with respect to $t$.
For example, for a given Fourier mode with energy $E$ in the observer frame, propagating along the $z$-direction, the driving force matrix reads 
\begin{align}\label{standard}
S_{ij}=\frac{E^2}{2} \left(
\begin{array}{ccc}
h_+&h_{\times}&0\\
h_{\times}&-h_+&0\\
0&0&0
\end{array}
\right)\,,
\end{align}
where $h_+$ and $h_\times$ denote the two polarisations of the GW.

The signal observed in an interferometer (called the strain and denoted by $h$) is directly proportional to the difference in length between its two arms. This is obtained by integrating twice Eq.~\eqref{geo2} to find the change in length induced by the passing of the GW. For an interferometer with arms pointing in direction $\hat{\mathbf{l}}$ and $\hat{\mathbf{m}}$,
in the long-wavelength regime,\footnote{For simplicity we assume that the long-wavelength approximation is valid, i.e.\ we ignore the effect of finite travel time of the photon.
We note that this will not be acceptable for LISA and therefore one would need to properly account for the time delays in the response function. It is worth noting that an extra complication would be the much longer observation time at the mHz frequencies where the impact of the source's velocity and acceleration would be even more important.} one finds that the strain is given by
\begin{equation}
h=\frac{1}{2}(\hat{l}_i\hat{l}_j-\hat{m}_i\hat{m}_i)P_{ij}\, ,  \label{eq:strain_def}  
\end{equation}
where 
\begin{align}\label{eq:PijfromSij}
P_{ij} \equiv \frac{2}{E^2} S_{ij}\, .
\end{align}
The $E^2$ factor  comes from the double integration over time when solving for $\xi^i$. We see that the dimensionless driving force matrix, $P_{ij}$, is the quantity that directly drives the amplitude of the detected GW signal. Note that for a monochromatic wave, the dimensionless driving force matrix is directly given by the metric in TT gauge: $P_{ij}=h^{\rm TT}_{ij}$. 

\section{Moving sources}\label{theory}

Let us now generalise these results to the case in which the source has a non-vanishing  peculiar velocity with respect to the observer frame. In this case, we need to distinguish between quantities calculated in the observer frame, and quantities calculated in a frame comoving with the source  (hereafter ``source frame''). Quantities in the source frame are denoted with a tilde. We denote by $\bv$ the velocity of the source with respect to the observer. 

\subsection{The dimensionless driving force matrix}

We start by calculating the dimensionless driving force matrix. In the source frame, we write the metric in the TT gauge, i.e.\ we have $\tilde{h}_{00}=\tilde{h}_{0i}=0$. The metric in the observer frame is related to the one in the source frame by a boost transformation $\Lambda$ with velocity $-\mathbf{v}$:
\begin{align}\label{LLtransformation}
h_{\mu\nu}=\Lambda_{\mu}^{\ \alpha}\Lambda_{\nu}^{\ \beta}\tilde{h}_{\alpha\beta} \, . 
\end{align}
Keeping only terms that are linear in the velocity, we obtain (see Appendix~\ref{BoostH} for details)
\begin{subequations}
\label{boost}
\begin{align}
h_{00}&=0 \,,\\
h_{0i}&=v_m \widetilde{h}_{mi}\,,\\
h_{ij}&=\widetilde{h}_{ij}\,. 
\end{align}
\end{subequations}

The geodesic deviation equation in the observer frame is still given by Eq.~\eqref{geo},  but the driving force matrix~\eqref{Relectric} has now a different form, due to the velocity of the source. Inserting Eqs.~\eqref{boost} into Eq.~\eqref{eq:driving} we obtain
\be
S_{ij}=\frac{1}{2}\left(-\partial_0^2\widetilde{h}_{ij}+v_m\partial_i\partial_0 \widetilde{h}_{mj}+v_m\partial_j \partial_0\widetilde{h}_{mi}\right)\,, \label{eq:Svel}
\ee
where $\partial_\mu$ denotes a derivative with respect to the spatial coordinate $x^\mu$. Let us assume that $\tilde h$ is a plane wave in the source frame. For a monochromatic wave we have:
\begin{align}
\widetilde{h}_{ij}
=\mathcal{A}_{ij}(\tilde{\mathbf{k}})
\exp\big(i\tilde{k}_\mu\tilde{x}^\mu\big)
+\mathcal{A}^{*}_{ij}(\tilde{\mathbf{k}})
\exp\big(-i\tilde{k}_\mu\tilde{x}^\mu\big)\,, \label{eq:tildeh}
\end{align}
where $\tilde{k}^\mu=(-\tilde E,\tilde{k}^i)$ and $\tilde{x}^\mu=(-\tilde\tau,\tilde x^i)$. 
Since $\tilde k_\mu \tilde x^\mu = k_\mu x^\mu$, the partial derivatives are handled using that $\partial_\mu \partial_\nu$ brings a factor $-k_\mu k_\nu$. With $E\equiv k_0$ the energy in the observer frame, we get
\begin{subequations}
\begin{align}
&\partial_0^2\widetilde{h}_{ij}= -E^2 \widetilde{h}_{ij}\, ,\\
&v_m\partial_0\partial_i\widetilde{h}_{mj}=-v_mk_i E\,\widetilde{h}_{ij}\, ,
\end{align}
\end{subequations}
leading to
\be
S_{ij}=\frac{1}{2}\left(E^2 \widetilde{h}_{ij}-E v_m k_i \widetilde{h}_{mj}-E v_m k_j \widetilde{h}_{mi}\right)\,. \label{eq:driving_vel}
\ee
The geodesic deviation equation in the observer frame is therefore directly affected by the peculiar velocity of the source. Note that in Eq.~\eqref{eq:driving_vel}, $E$ is shifted with respect to the energy in the source frame, $\tilde E$, by the velocity along the direction of propagation
\begin{align}
E=\tilde{E}\big(1+\bv\cdot\tbn \big)\, ,
\end{align}
where $\tbn$ is a unit vector along the direction of propagation: $\tbn\equiv \tilde{\bk}/\tilde{k}$. 

As before, the geodesic deviation equation must be solved to find the length difference between the two arms of an interferometer. Since $\tilde{h}_{\mu\nu}$ in Eq.~\eqref{eq:driving_vel} depends on the proper time of the source, $\tilde\tau$, we first rewrite Eq.~\eqref{geo} in terms of $\tilde\tau$ using that $dt=d\tau$ and that $d\tau=\tilde{E}/E \,d\tilde\tau$ (see Appendix~\ref{app:dtau} for a derivation of this equality). We obtain
\begin{align}
\label{geotilde}
\frac{d^2\xi^i}{d\tilde{\tau}^2}=- \left(\frac{\tilde{E}}{E}\right)^2S_{ij} \xi^j\,.
\end{align}
Using Eqs.~\eqref{eq:driving_vel} and~\eqref{eq:tildeh}, and integrating twice over proper time, we find that the dimensionless driving force matrix is given by
\begin{align}
P_{ij}=\frac{2}{\tilde{E}^2} \left(\frac{\tilde{E}}{E}\right)^2S_{ij} =  \frac{2}{E^2} S_{ij}\, . \label{eq:Pijvel}
\end{align}
We see that the $1/\tilde{E}^2$ factor coming from the integration of $\tilde{h}_{\mu\nu}$ in Eq.~\eqref{eq:tildeh} over proper time $\tilde{\tau}$ cancels the $\tilde{E}^2$ factor in Eq.~\eqref{geotilde}. Note that the same result can be found by rewriting $\tilde{h}_{\mu\nu}$ in terms of quantities in the observer frame, using that $\tilde k_\mu \tilde x^\mu = k_\mu x^\mu$, and then integrating Eq.~\eqref{geo} directly over the proper time of the observer. Here for simplicity we have derived Eq.~\eqref{eq:Pijvel} for a monochromatic plane wave. However, using the second method, it is easy to show that the same dimensionless driving form matrix $P_{ij}$ is found for any waveform that can be written as a superposition of plane waves with same direction $\tbn$ but different energies $\tilde{E}$.

As an example, let us compute $P_{ij}$ for a wave propagating along the $z$-direction (in the source frame), i.e.\ $\tilde{\bn}=(0,0,1)$.  Using that at zeroth order in the velocity $k_i= E n_i= E\tilde{n}_i$, leading to $E v_m k_i=E^2 v_m\tilde{n}_i$, and inserting Eq.~\eqref{eq:driving_vel} into~\eqref{eq:Pijvel} we obtain
\begin{align}\label{eq:Sij_vel}
P_{ij}= \left(
\begin{array}{ccc}
\thh_+&\thh_{\times}&\!\!\!\!-v_x\thh_+ -v_y\thh_{\times}\\
\thh_{\times}&-\thh_+&\!\!\!\!-v_x\thh_{\times}+v_y \thh_+\\
\!\!\!\!-v_x\thh_+-v_y \thh_{\times}&\!\!\!\!-v_x \thh_{\times}+v_y \thh_+&0
\end{array}
\right)
\end{align}
where $\thh_+$ and $\thh_\times$ are the plus and cross polarisations in the source frame. Comparing Eq.~\eqref{eq:Sij_vel} with Eq.~\eqref{standard}, we see that the relative motion of the source with respect to the observer generates contributions to the dimensionless driving force matrix that are not transverse to the GW direction $\tilde{\bn}$. $P_{ij}$ in Eq.~\eqref{eq:Sij_vel} has indeed non-zero contributions in direction $zx$ and $zy$. In the next section we determine the observable impact of these non-transverse contributions.

In general, for a wave propagating in arbitrary direction, we define a set of orthonormal vectors, adapted to the incoming direction of the wave in the source frame
\begin{subequations}\label{eq:tall}
\begin{align} 
\tilde\bn &= (\sin\ttheta \cos\tphi, \sin\ttheta\sin\tphi, \cos\ttheta)\, , \label{eq:tn}\\
\tilde{\mathbf{e}}_1 (\tbn) &= (\sin\tphi, -\cos\tphi, 0)\, , \label{eq:te1}\\
\tilde{\mathbf{e}}_2 (\tbn) &= (\cos\ttheta \cos\tphi, \cos\ttheta\sin\tphi, -\sin\ttheta)\, .\label{eq:te2}
\end{align}
\end{subequations}
With respect to these vectors, the metric in the TT gauge can be decomposed as
\begin{align}
\tilde{h}_{ij}^{\rm TT}=\tilde h_+ \big(\te_{1i} \te_{1j} - \te_{2i}\te_{2j}\big) + \tilde h_\times\big(\te_{1i}\te_{2j}+\te_{2i}\te_{1j}\big)\,,
\end{align}
where it is implied that $\tilde h_{+,\times} = \tilde h_{+,\times}(\tbn)$.
Inserting this into Eqs.~\eqref{eq:driving_vel} and~\eqref{eq:Pijvel}, and using as before that at linear order in the velocity $Ev_mk_i=E^2 v_m\tilde{n}_i$, we obtain
\begin{align}
P_{ij} =& h_+ \big(\te_{1i} \te_{1j} - \te_{2i}\te_{2j}\big) + h_\times\big(\te_{1i}\te_{2j}+\te_{2i}\te_{1j}\big) \nn\\
&+ h_1 \big(\tn_i \te_{1j} + \te_{1i} \tn_j\big) + h_2 \big(\tn_i \te_{2j} + \te_{2i} \tn_j\big)\,, \label{eq:hijGen}
\end{align}
where
\begin{align}\label{master}
h_+ &= \thh_+\, ,\\
h_\times &=  \thh_\times\, ,\nn\\
h_1 &= -v_1 \thh_+ - v_2 \thh_\times\, ,\nn\\
h_2 &= -v_1 \thh_\times + v_2 \thh_+\, ,\nn
\end{align}
and we have defined the velocity component along the orthonormal set
\begin{align}
v_1\equiv \bv\cdot\tilde{\mathbf{e}}_1\,, \quad v_2\equiv \bv\cdot\tilde{\mathbf{e}}_2\,, \quad v_3\equiv \bv\cdot\tbn\, .   
\end{align}
As before, we see that the source velocity generates contributions to $P_{ij}$ that are longitudinal: $h_1$ and $h_2$ are indeed along the direction of propagation $\tilde{\bn}$.

Before moving to the calculation of the strain, let us comment on the relation between the dimensionless driving force matrix and the metric in the TT gauge. In the case of non-moving sources we saw that the dimensionless driving force matrix is equal to the metric in the TT gauge. For a moving source we note that the symmetry between source and observer reference frames is broken. Hence fixing the TT gauge in one frame is no longer preserved under transformation on to the other frame. The dimensionless driving force matrix $P_{ij}$ is therefore no longer equal to the metric in TT gauge \emph{in the source frame}. However, we can apply another gauge transformation to the metric $h_{\mu\nu}$, to bring it in the TT gauge in the observer frame. In that case, we show in Appendix~\ref{Gauge} that the resulting metric~\eqref{eq:hTTobs} becomes equal to the dimensionless driving form matrix~\eqref{eq:Sij_vel}.

\subsection{The strain}
\label{sec:pattern}

We now project the dimensionless driving force matrix $P_{ij}$ onto the arms of an interferometer $\hat{\mathbf{l}}$ and $\hat{\mathbf{m}}$ to obtain the strain
\begin{align}
h&=\frac{1}{2}(\hat{l}_i \hat{l}_j - \hat{m}_i \hat{m}_j)P_{ij}  \\
&=F_+(\tbn) h_+ + F_\times(\tbn) h_\times\label{eq:strain_spin1}+ F_1(\tbn) h_1 + F_2(\tbn) h_2 \,,\nn
\end{align}
where the antenna patterns are given by
\begin{align}
F_+ (\tbn) &= \frac{1}{2}\big(\hat{l}_i \hat{l}_j - \hat{m}_i \hat{m}_j\big)  \big(\te_{1i} \te_{1j} - \te_{2i}\te_{2j}\big)\, , \nonumber \\
F_\times (\tbn) &= \frac{1}{2}\big(\hat{l}_i \hat{l}_j - \hat{m}_i \hat{m}_j\big)  \big(\te_{1i}\te_{2j}+\te_{2i}\te_{1j}\big)\, , \nonumber \\
F_1 (\tbn) &= \frac{1}{2}\big(\hat{l}_i \hat{l}_j - \hat{m}_i \hat{m}_j\big)  \big(\tn_i \te_{1j} + \te_{1i} \tn_j\big)\, , \nonumber \\
F_2 (\tbn) &= \frac{1}{2}\big(\hat{l}_i \hat{l}_j - \hat{m}_i \hat{m}_j\big)  \big(\tn_i \te_{2j} + \te_{2i} \tn_j\big) \, .
\end{align}

As an example let us consider the strain response of an interferometer with arms pointing in the $x$ and $y$ directions: $\hat{\mathbf{l}}=(1,0,0)$ and $\hat{\mathbf{m}}=(0,1,0)$. We obtain
\begin{align}
h&=\frac{1}{2}(\hat{l}_i\hat{l}_j-\hat{m}_i\hat{m}_j)P_{ij}=-\frac{\thh_+}{2}\left(\cos^2\ttheta+1\right)\cos2\tphi\nn\\
&+\thh_\times\cos\ttheta\sin2\tphi -\left(v_1 \thh_+ + v_2 \thh_\times\right)\sin\ttheta\sin2\tphi\nonumber\\
&-\left(v_1 \thh_\times- v_2 \thh_+\right)\sin\ttheta\cos\ttheta\cos2\tphi\,. \label{eq:strainxy}
\end{align}
From Eq.~\eqref{eq:strainxy} we see that the transverse velocity of the source, namely the components $v_1$ and $v_2$, generates contributions to the signal \emph{which are not} proportional to the spin-2 antenna patterns
\begin{subequations}
\begin{align}
&F_+(\tbn)=-\frac{1}{2}(\cos^2\ttheta+1)\cos 2\tphi\,, \\ 
&F_\times(\tbn)=\cos\ttheta\sin2\tphi\, .
\end{align}
\end{subequations}
These new contributions are proportional instead to spin-1 antenna patterns $F_1$ and $F_2$. 

In Eqs.~\eqref{eq:strain_spin1} and~\eqref{eq:strainxy} we have identified spin-2 modes as the contributions that are transverse to the direction of propagation of the GW, $\tilde{\bn}$, in the source frame. This definition is somewhat arbitrary, since we do not observe $\tilde{\bn}$ directly: we reconstruct it from the antenna patterns $F_+(\tilde{\bn})$ and $F_\times(\tilde{\bn})$. We can therefore wonder if there exists a direction $\bn$ such that the strain would contain only spin-2 polarisations \emph{with respect to that direction}. In the next section we show that this is indeed the case, and that this new direction $\bn$ is nothing else than the aberrated direction obtained by applying the boost transformation on $\tilde{k}^{\mu}$ (and extracting the spatial part of the resulting vector).

\section{Aberrated reference frame}
\label{sec:aberrated}

As for electromagnetic signals, we can define an aberrated momentum $k^\mu$ by applying the boost $\Lambda^\mu_{\ \nu}$ on $\tilde{k}^{\mu}$. The spatial part of $k^\mu$ is given by
\begin{align}
k^i=\Lambda^i_{\ \mu}\tilde{k}^\mu=\tilde{E}\left( \tilde{n}^i+v^i\right)\, ,
\end{align}
leading to
\begin{align}
\bn\equiv\frac{\mathbf{k}}{|\mathbf{k}|}=  \tbn+\mathbf{v}-v_3\tbn= \tbn+\bv_\perp\, ,\label{eq:relation_n}
\end{align}
where the transverse velocity $\bv_\perp$ is defined as
\begin{align}
\bv_\perp=\bv-v_3\tbn\, .
\end{align}
Note that this velocity is transverse to both $\bn$ and $\tbn$ since we neglect contributions quadratic in the velocity.

Let us start by calculating the strain for a detector with arms along $xy$, given by Eq.~\eqref{eq:strainxy}. From Eq.~\eqref{eq:relation_n} we find that the aberrated angles are related to angles at the source by
\begin{subequations}\label{eq:ab}
\begin{align}
\theta&=\ttheta+\delta\theta=\ttheta+v_2\, \label{eq:ab0} ,\\
\phi&=\tphi+\delta\phi=\tphi-\frac{v_1}{\sin\theta}\, .\label{eq:ab1}
\end{align}
\end{subequations}
The apparent divergence at $\theta =0$ is an artefact of the coordinate singularity there. The right ascension $\phi$  is indeed ambiguous at $\theta = 0$.
Inserting this into Eq.~\eqref{eq:strainxy} we obtain for the strain
\begin{align}\label{eq:strainxy_aber}
h=\frac{1}{2}(\hat{l}_i\hat{l}_j-\hat{m}_i\hat{m}_j)&P_{ij}=\thh_+\left[ F_+(\bn)+2v_1\frac{\cos\theta}{\sin\theta}F_{\times}(\bn)\right]\nn\\
+&\thh_\times\left[ F_\times(\bn)-2v_1\frac{\cos\theta}{\sin\theta}F_{+}(\bn)\right]
\,. 
\end{align}
We see that the source velocity induces a mixing between the two polarisations, proportional to $F_+$ and $F_{\times}$. Defining the polarisation angle
\begin{equation}\label{mixing}
\delta\psi=-v_1\frac{\cos\theta}{\sin\theta}\,,    
\end{equation}
we can rewrite Eq.~\eqref{eq:strainxy_aber} as
\begin{align}
h=&\frac{1}{2}(\hat{l}_i\hat{l}_j-\hat{m}_i\hat{m}_j)P_{ij}\nonumber\\
=&\,\thh_+\Big[ F_+(\bn)\cos(2\delta\psi)-F_{\times}(\bn)\sin(2\delta\psi)\Big]\nn\\
&+\thh_\times\Big[ F_\times(\bn)\cos(2\delta\psi)+F_{+}(\bn)\sin(2\delta\psi)\Big] \nn\\
=&\,\hat{h}_+(\bn)F_+(\bn)+\hat{h}_\times(\bn) F_\times(\bn)\, ,\label{eq:strain_hat}
\end{align}
where we have defined
\begin{subequations}\label{eq:h_aberr}
\begin{align}
\hat{h}_+(\bn)&\equiv \thh_+(\tbn)\cos(2\delta\psi)+\thh_\times(\tbn) \sin(2\delta\psi) \, ,\label{eq:hplus_aberr}\\
\hat{h}_\times(\bn)&\equiv\thh_\times(\tbn)\cos(2\delta\psi)-\thh_+(\tbn) \sin(2\delta\psi) \, .\label{eq:hcross_aberr}
\end{align}
\end{subequations}
With respect to the aberrated direction, the strain contains therefore only spin-2 modes, proportional to the spin-2 antenna patterns $F_+$ and $F_\times$.
Parameter estimations from the GW signal will therefore infer: 1) the aberrated direction $\bn$ and 2) the two "mixed" polarisations $\hat{h}_+$ and $\hat{h}_\times$. Since the transverse peculiar velocity of the source is unknown, the mixing angle (\ref{mixing}) is unknown, and we can therefore not measure the two intrinsic polarisations $\thh_+$ and $\thh_\times$.

We could wonder if having detectors with arms pointing in different directions could help us break the degeneracy between the source velocity and the true polarisations. Eq.~\eqref{eq:strain_hat} has indeed been derived in the specific case of a detector with arms pointing in the $xy$ directions. We can show that the degeneracy exists for all cases. At linear order in the velocity, we can indeed rewrite Eqs.~\eqref{eq:hijGen} and~\eqref{master} as
\begin{align}
&P_{ij} =\label{eq:Pij_general}\\
& \Big[ (\te_{1i} - v_1 \tn_i) (\te_{1j} - v_1 \tn_j) - (\te_{2i} - v_2 \tn_i) (\te_{2j} - v_2 \tn_j) \Big]  \thh_+\nn\\
&+ \Big[(\te_{1i} - v_1 \tn_i) (\te_{2j} - v_2 \tn_j) + (\te_{1j} - v_1 \tn_j) (\te_{2i} - v_2 \tn_i) \Big]  \thh_\times\nn\, .
\end{align}
We see that, working to linear order in velocity, the boosted dimensionless driving force matrix is equivalent to the one of an unboosted gravitational wave with polarisation axes
\begin{subequations}\label{eq:eall}
\begin{align}
\mathbf{e}_1 &= \tilde{\mathbf{e}}_1 - v_1 \tbn\,,\label{eq:e1}\\ 
\mathbf{e}_2 &= \tilde{\mathbf{e}}_2 - v_2 \tbn\,.\label{eq:e2}
\end{align}
\end{subequations}
It is clear that these two polarisation vectors are orthogonal, and they correspond to the polarisation axes of a source coming from direction
\begin{align}
\bn &= \mathbf{e}_1 \wedge \mathbf{e}_2 = \tbn + v_1 \tilde{\mathbf{e}}_1 +v_2 \tilde{\mathbf{e}}_1=\tbn+\bv_\perp\, , \label{eq:kaberr}
\end{align}
which is nothing else than the aberrated direction defined in Eq.~\eqref{eq:relation_n}. The polarisation axes $\mathbf{e}_1$ and $\mathbf{e}_2$ are not the natural ones associated to the direction $\bn$, as defined in Eqs.~\eqref{eq:tall}. We can easily see that the natural axes are related to $\mathbf{e}_1$ and $\mathbf{e}_2$ by 
\begin{subequations}\label{eq:hat_eall}
\begin{align}
\hat{\mathbf{e}}_1 &= \mathbf{e}_1 - v_1\frac{\cos\theta}{\sin\theta} \mathbf{e}_2\,, \label{eq:hat_e1} \\
\hat{\mathbf{e}}_2 
&=\mathbf{e}_2 + v_1\frac{\cos\theta}{\sin\theta} \mathbf{e}_1\,,\label{eq:hat_e2}
\end{align}
\end{subequations}
This is the infinitesimal form of a rotation in two dimensions 
\begin{equation}\label{eq:DefRab}
\hat{\mathbf{e}}_a = R_a^{\,\,b}  {\mathbf{e}}_b\,,
\end{equation}
where the rotation matrix is
\begin{equation}
R_a^{\,\,b}  = \left(\begin{array}{cc}
\cos(\delta \psi)&\sin(\delta \psi)\\
-\sin(\delta \psi )&\cos(\delta \psi)
\end{array}\right) \simeq \left( \begin{array}{cc}
1&\delta \psi\\
-\delta \psi &1
\end{array}\right)\,,
\end{equation}
and $\delta \psi$ is defined in Eq.\,(\ref{mixing}).  
Inserting Eqs.~\eqref{eq:hat_eall} into Eq.~\eqref{eq:Pij_general} we obtain
\begin{align}
P_{ij} =\hat{h}_+ \big(\hat{e}_{1i}\hat{e}_{1j}-\hat{e}_{2i}\hat{e}_{2j}\big)+\hat{h}_\times\big(\hat{e}_{1i}\hat{e}_{2j}+\hat{e}_{2i}\hat{e}_{1j}\big)\, ,\label{eq:Pij_hat}
\end{align}
where $\hat{h}_+$ and $\hat{h}_\times$ are given by Eqs.~\eqref{eq:h_aberr}. From this we see that the response of \emph{any} interferometer can be written in terms of the two standard antenna patterns $F_+(\bn)$ and $F_\times(\bn)$ associated to the aberrated direction $\bn$. The two inferred polarisations $\hat{h}_+$ and $\hat{h}_\times$ are modified by the source velocity. 

Eq.~\eqref{eq:Pij_hat} tells us that the spin-1 modes that are generated by the velocity of the source can always be re-absorbed into spin-2 modes with aberrated direction $\bn$ and mixed polarizations $\hat{h}_+$ and $\hat{h}_\times$. One could however wonder if by actively searching for vector modes, i.e.\ by including spin-1 antenna patterns in the modelling of the signal, one could measure the amplitude of these new modes, as well as the true direction $\tbn$. This turns out to be impossible, since there is no unique way of splitting the signal into spin-2 modes and spin-1 modes, see Appendix \ref{vec} for details. 

\begin{figure}
    \centering
    \includegraphics[scale=0.5]{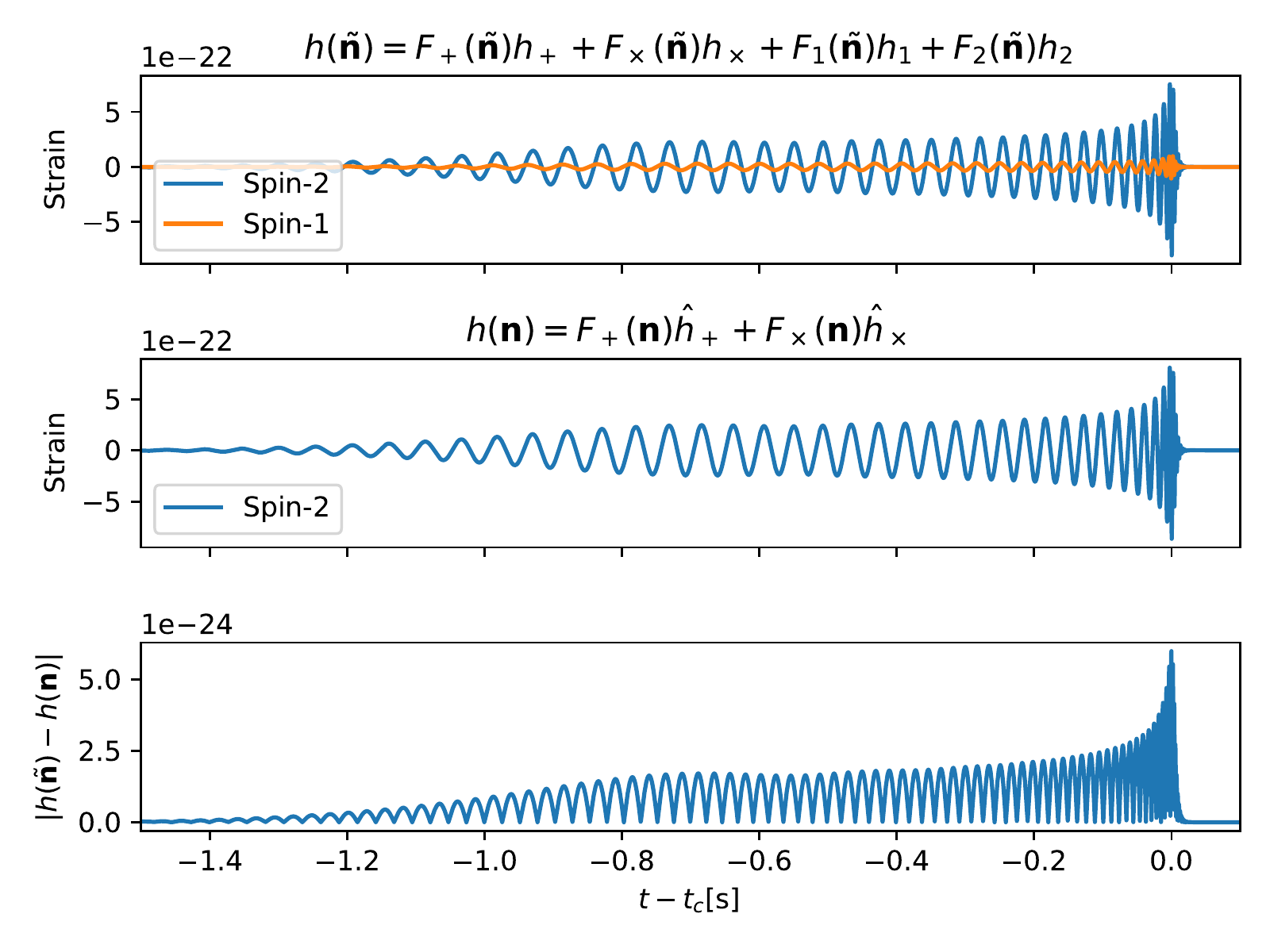}
    \caption{The first and second panels show two templates constructed for a GW signal from a binary with $30 M_\odot-30 M_\odot$ at a distance of 500 Mpc and with $\mathbf{v}=(0.1,0,0)$. In the first panel, we construct the template as the sum of two spin-1 (orange line) and spin-2 signals (blue line) using the true sky direction $\tilde{\mathbf{n}}$ and the true polarisation angle $\tilde{\psi}$. In the second panel, we construct the template using the aberrated sky position $\mathbf{n}$ and aberrated polarisation angle $\psi$ and only spin-2 modes. The third panel shows the difference between the two templates which is of the order of $\mathcal{O}(\mathbf{|v|}^2)$, as our framework is defined at the first order in peculiar motion. The detector is taken with arms $\hat{\mathbf{l}}=(1,0,0)$ and $\hat{\mathbf{m}}=(0,1,0)$.
    }
    \label{fig:pecvel}
\end{figure}

As a consequence, from a data analysis point of view, the template of a signal constructed for the direction of propagation $\tilde{\mathbf{n}}$ using spin-2 modes, spin-1 modes and the polarisation angle $\tilde{\Psi}$ will be equivalent (up to a factor $\mathcal{O}(|\mathbf{v}|^2)$) to the template of a signal propagating in the direction  $\mathbf{n}$ with spin-2 modes and polarisation angle $\Psi$. Fig.~\ref{fig:pecvel} shows the templates for the true and the aberrated directions of a simulated GW signal with $|\mathbf{v}|=0.1$ (see caption for more details). For the true direction, the template is the sum of spin-2 and spin-1 components. As expected, the two templates differ by a factor $\mathcal{O}(|\mathbf{v}|^2)$.

Another manner to understand this total degeneracy is to consider the geometric interpretation of the transformation of the dimensionless driving force matrix from the source frame to the observer frame. Let us consider the rotation vector $\mathbf{\cal A}_{\tilde{n} \to n} = \alpha (\tbn \wedge \hat{\bv}_\perp)$, where $\alpha$ is the angle between $\tbn$ and $\bn$ and $(\tbn \wedge \hat{\bv}_\perp)$ is a unit vector orthogonal to them. The rotation around $\mathbf{\cal A}_{\tilde{n} \to n}$ carries $\tbn$ along a great circle to $\bn$. Its components are
\begin{equation}
R_i^{\,\,j} \equiv \text{exp}\left(- {\cal A}_{\tilde{n} \to n}^k \epsilon_{ki}^{\phantom{ki}j}\right) \simeq \delta_i^j - \tilde{n}_i v^j + v_i \tilde{n}^j \,,\label{eq:Rij}
\end{equation}
and the transformation rules \eqref{eq:eall} are directly seen as the effect of this latter infinitesimal rotation since they are equivalent to $e_{1i} = R_i^{\,\,j}\tilde{e}_{1j}
$ and $e_{2i} = R_i^{\,\,j}\tilde{e}_{2j}$. Therefore Eq.~\eqref{eq:Pij_general} is simply 
\begin{equation}\label{eq:PRRPtilde}
P_{ij}(\bn) = R_i^{\,\,p} R_j^{\,\,q} \tilde P_{pq}(\tbn)\, ,
\end{equation}
with $\tilde{P}_{ij}=\tilde{h}_{ij}^{\rm TT}$. We recognize the transformation rule of a tensor on the unit sphere under a rotation $R$. Hence the driving force matrix is also transformed by the infinitesimal rotation which transports $\tbn$ onto $\bn$. Note that this transformation is equivalent to a parallel transport of the driving force along the great circle connecting $\tbn$ and $\bn$, as by construction both vectors lie in the equatorial plane of vectors normal to $\mathbf{\cal A}_{\tilde{n} \to n}$.

However, even though in the source frame we chose for convenience to use the vectors naturally associated with the spherical components (Eqs. \eqref{eq:te1} and \eqref{eq:te2}), the rotated ones, $\mathbf{e}_1$ and $\mathbf{e}_2$, are not directly the unit vectors naturally associated with spherical coordinates in the observer frame: $\hat{\mathbf{e}}_1$ and $\hat{\mathbf{e}}_2$. Both sets being orthonormal and normal to $\bn$, they are related by a rotation around $\bn$ of angle $\delta \psi$, that is Eq.~\eqref{eq:DefRab}, with $R_a^{\,\,b}$ related to $R_i^{\,\,j}$ through $R_a^{\,\,b} \equiv  {\hat{e}_a}^i {e}_{bi}  = {\hat{e}_a}^i R_i^{\,\,j}\tilde{e}_{bj}$.

Therefore, from the simple transformation rule~\eqref{eq:PRRPtilde}, the spherical basis components of the driving force, which are $\tilde{P}_{ab} \equiv {\tilde{e}_a}^i {\tilde{e}_b}^j \tilde{P}_{ij}$ and $P_{ab} \equiv {\hat{e}_a}^i {\hat{e}_b}^j P_{ij}$, are related through
\begin{equation}\label{eq:RotationRab}
P_{ab}(\bn) = R_a^{\,\,c} R_b^{\,\,d}  \tilde{P}_{cd}(\tbn)\,.
\end{equation}
This is yet another way to write the transformation rule of a tensor on the unit sphere under a rotation, which translates into Eqs.~\eqref{eq:h_aberr} for the polarisation components. In short, the mixing of polarisations is essentially a consequence of the fact that the basis used to define polarisations, the natural spherical basis, is not parallel transported along the great circle connecting $\tbn$ to $\bn$, whereas the driving force matrix is parallel transported. The only exception is when the great circle connecting $\tbn$ to $\bn$ is either the equator or a meridian of the spherical coordinates system. For infinitesimal transformations which we have considered here, the natural spherical basis is also (infinitesimally) parallel transported whenever the direction (initial or final, this is equivalent for infinitesimal transformations) is on the equator, even if the transformation direction is not tangential to the equator. That is whenever the conditions $\theta = \pi/2$ (emitting direction on the equator) or $v_1=0$ (an aberration along a meridian) are satisfied, the natural spherical basis is infinitesimally parallel transported, and we can check that indeed $\delta \psi =0$ under these conditions. 

 Also, one should bear in mind that the rotation \eqref{eq:Rij} which accounts for the effect of the transverse velocity depends on $\tbn$ and is not a unique global rotation. Therefore, a source with a transverse velocity is degenerate from a source without velocity but rotated with $R$, only because we can observe a single emission direction. Finally, let us highlight that the transformation of the driving force matrix due to a transverse velocity, seen as rotation or as parallel transport, is similar to the transformation of the CMB polarization tensor which is also a spin-2 quantity, see e.g. Section III of \citet{Challinor:2002zh}.

\section{Time delay from a network of detectors}
\label{sec:time delay}

\begin{figure}
    \centering
    \includegraphics[width=0.2\textwidth]{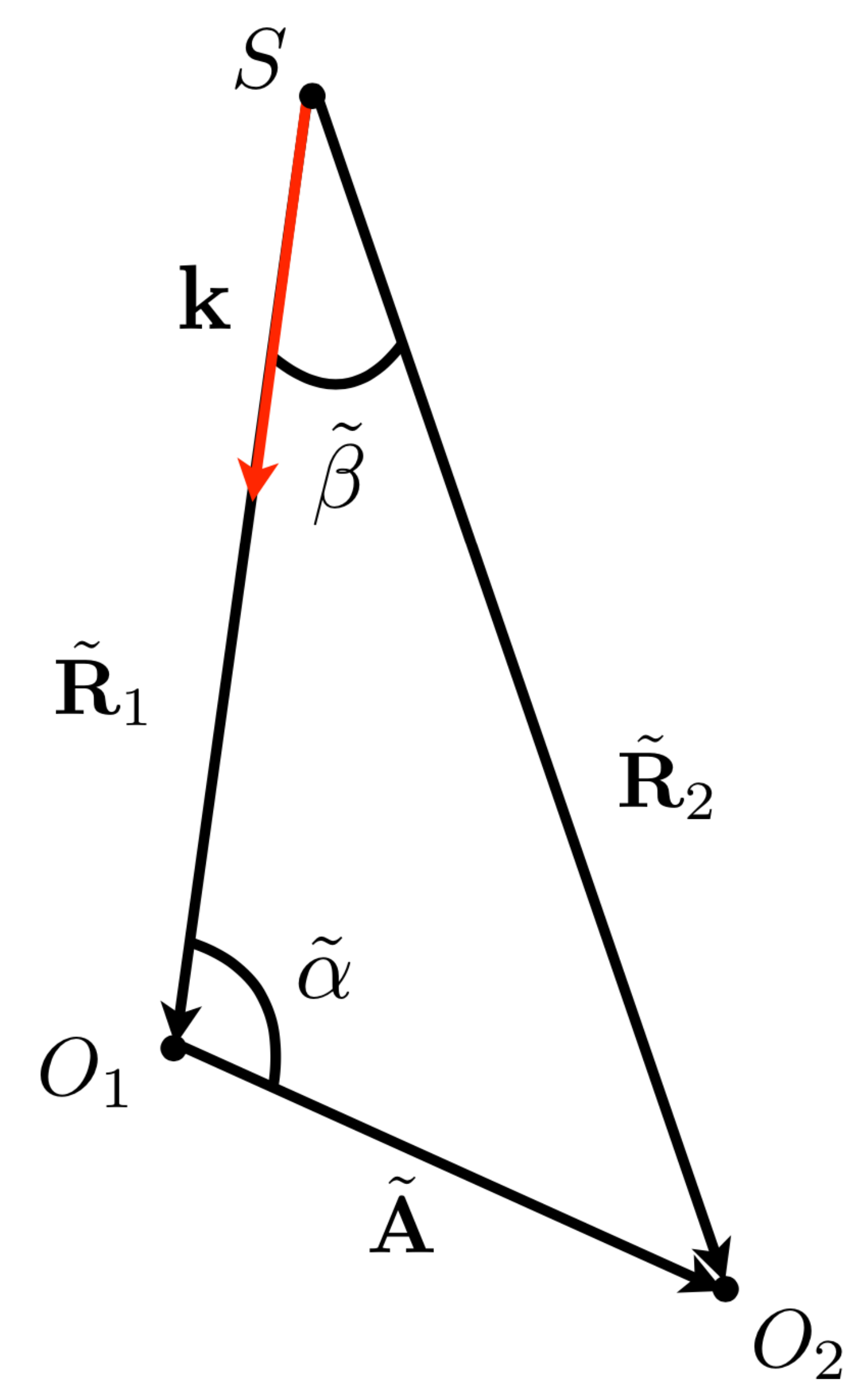}
    \caption{Geometrical configuration used to calculate the time delay. }
    \label{fig:time_delay}
\end{figure}

For a network of interferometers, in addition to the signal measured by each detector, the time delay between the different detectors due to their different position with respect to the source is measured. This time delay depends directly on the direction of the source, and provides therefore a precise way of measuring this direction (more precise than from the antenna patterns, since the phase of the GW is measured with a better precision than the amplitude of the two polarisations). We will see that in the case of a moving source the time delay depends on the aberrated direction $\bn$ and not on the true direction $\tbn$.

We consider the geometry plotted in Fig.~\ref{fig:time_delay}. In the reference frame of the source (denoted by tilde), the source emits a GW at time $\tilde t_e=0$ and at position $\tilde{\mathbf{R}}_e=(0,0,0)$. The first interferometer receives the wave at time $\tilde t_1$ and position $\tilde{\mathbf{R}}_1$, where $\tilde{R}_1=\tilde{t}_1$ (let us recall that we work in units $c=1$). The observer, who is moving with a velocity $-\mathbf{v}$ with respect to the source, sees boosted coordinates $x_\mu=(t,\mathbf{R})=\Lambda_\mu^{\ \nu}\tilde{x}_\nu$, where $\Lambda$ is defined in~Appendix~\ref{BoostH}. At linear order in the velocity the time of emission and reception are given by
\begin{subequations}
\begin{align}
 t_e&=\tilde{t}_e+\mathbf{v}\cdot\tilde{\mathbf{R}}_e=0\, ,  \\
 t_1&=\tilde{t}_1+\mathbf{v}\cdot\tilde{\mathbf{R}}_1=\tilde{R}_1+\mathbf{v}\cdot\tilde{\mathbf{R}}_1\, .
\end{align}
\end{subequations}
The same calculation applies to the second interferometer. The difference in arrival time between the two detectors is therefore given by
\be
\Delta t\equiv t_2-t_1=\tilde{R}_2-\tilde{R}_1+\mathbf{v}\cdot(\tilde{\mathbf{R}}_2-\tilde{\mathbf{R}}_1)\, .
\ee
Defining $\tilde{\mathbf{A}}$ as the vector connecting the two detectors:
\be
\tilde{\mathbf{A}}=\tilde{\mathbf{R}}_2-\tilde{\mathbf{R}}_1\, ,
\ee
we see from Fig.~\ref{fig:time_delay} that 
\begin{subequations}
\begin{align}
&\tA^2=\tR_1^2+\tR_2^2-2\tR_1\tR_2\cos\tilde{\beta}\, ,\\   
&\tR_2^2=\tR_1^2+\tA^2-2\tR_1\tA\cos\tilde{\beta}\, ,
\end{align}
\end{subequations}
leading to
\be
\tR_2\cos\tilde\beta-\tR_1=-\tA\cos\tilde\alpha=\tbA\cdot\tbn\, .
\ee
We are interested in situations where the distance to the source is much larger than the distance between the detectors, such that $\cos\tilde{\beta}\simeq 1$. The time delay becomes then
\be
\Delta t=\tbA\cdot\left(\tilde{\mathbf{n}}+\mathbf{v}\right)\, .\label{eq:dtAtilde}
\ee
The distance between the two detectors in the source frame, $\tilde{\mathbf{A}}$, can be related to the distance in the observer frame using that 
\begin{align}
\tilde{R}_{1i}=\big(\Lambda^{-1}\big)_i^{\ \mu} x_{1\mu} =-v_i t_1 +R_{1i}\, ,  
\end{align}
and similarly for $\tilde{R}_{2i}$. This leads to
\begin{align}
\tilde{\mathbf{A}}=\mathbf{A}-\mathbf{v}\cdot \Delta t \, .
\end{align}
Inserting this in Eq.~\eqref{eq:dtAtilde} and keeping only terms at linear order in the velocity we obtain
\begin{align}
\Delta t=\bbA\cdot\left(\tbn+\mathbf{v}-v_3\tbn \right) =\bbA\cdot \left(\tbn+\mathbf{v}_\perp \right)=\bbA\cdot\bn\, . \label{eq:timedelay_aberr}
\end{align}
The time delay is therefore proportional to the aberrated direction $\bn$.

In practice, one often measures the phase shift between the waveform detected by two detectors at a fixed reference time, rather than the time delay. We can easily show that the phase shift is affected in the same way as the time delay by the source velocity. The phases at time $t$ and positions $\mathbf{R}_1$ and $\mathbf{R}_2$ are given by
\begin{subequations}
\begin{align}
\Phi(t,\mathbf{R}_1) &=-k^\mu x_{\mu 1}=E\left(t-\mathbf{R}_1\cdot\bn \right)\, , \\
\Phi(t,\mathbf{R}_2) &=-k^\mu x_{\mu 2}=E\left(t-\mathbf{R}_2\cdot\bn \right)\, ,
\end{align}
\end{subequations}
where $k^\mu=E(-1,\bn)$. The phase shift is given by
\begin{align}
\Delta\Phi=-E\, \bbA\cdot\bn \, .   
\end{align}
As expected, the phase shift is therefore also proportional to the aberrated direction $\bn$.

This calculation of the time delay (and the phase shift) shows that a network of detectors also measures the aberrated direction and not the intrinsic one in the source frame.

\section{Observational impact of the source velocity}\label{sec:obs}

\begin{figure}
    \centering
    \includegraphics[width=0.18\textwidth]{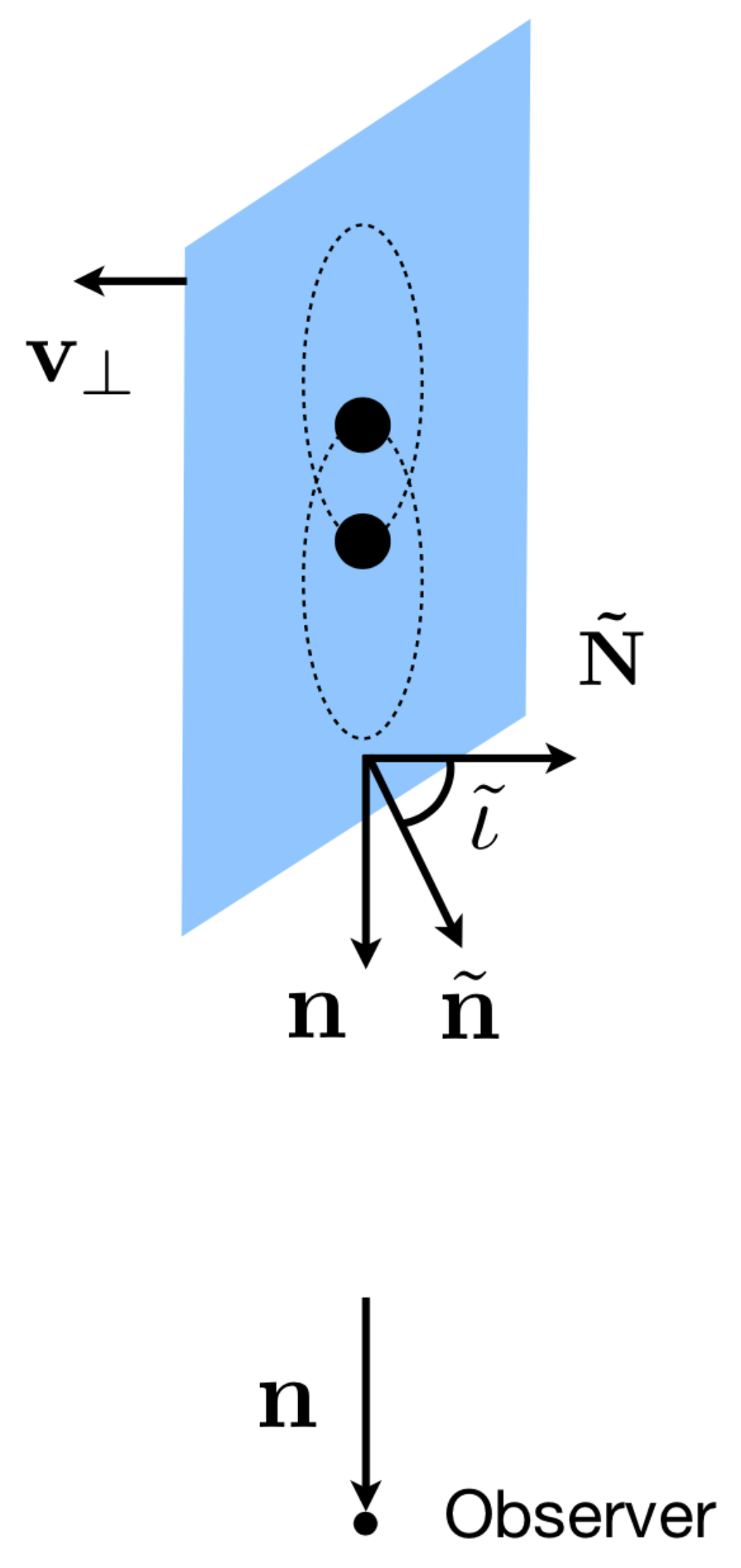}
    \caption{Sketch of the effect of aberration for a binary which is edge-on with respect to the observer. }
    \label{fig:iota}
\end{figure}

We have seen that the source velocity affects the strain in two ways: 1) it aberrates the direction of the source, both in the antenna patterns and in the time delay; and 2) it mixes the two polarisations of the wave. The first effect is common to any signal emitted by a moving source. In particular it affects in the exact same way electromagnetic signals. The second effect on the other hand is specific to the fact that a GW is a spinned quantity. This effect is therefore absent in standard optical or radio surveys, where we measure the intensity (which is a spin-zero quantity) of the electromagnetic field.\footnote{It would however be present if we were to measure directly the electromagnetic field, which is a spin-1 quantity.} These two effects have a direct impact on the measurement of the parameters of the binary.

As for electromagnetic signals, aberration means that we do not receive the GW that have been emitted in the observed direction $\bn$, but rather the GW that have been emitted in a different direction $\tbn$. As depicted in Fig.~\ref{fig:iota}, the source appears therefore in the correct position, but the form of the wave corresponds to the one emitted in direction $\tbn$. Since GW are not isotropically emitted by the binary system, aberration has a direct impact on the amplitude of the detected signal. In particular, even though the signal seems to come from direction $\bn$, the inclination angle that governs the amplitude of the signal is the one associated to the direction $\tbn$. The relation between $\tilde\iota$, defined as the angle between $\tbn$ and the normal to the plane of the binary in the frame of the source $\mathbf{\tilde{N}}$ (see Fig.~\ref{fig:iota}), and the true inclination angle $\iota$ that we would have if there would be no velocity (i.e.\ the angle between $\bn$ and $\mathbf{\tilde{N}}$) directly follows from the relation between $\bn$ and $\tbn$ and is therefore linear in the transverse velocity $\bv_\perp$. The polarisations $\tilde{h}_+$ and $\tilde{h}_\times$ in Eqs.~\eqref{eq:h_aberr} depend however not directly on $\tilde\iota$ but on its cosine, which is related to the one in absence of velocity by   
\begin{align}\label{Effect:iota}
\cos\tilde{\iota}=\cos\iota-\mathbf{\tilde{N}}\cdot\bv_\perp\, .
\end{align}
We see that the effect vanishes for a binary that is face-on, since in this case $\mathbf{\tilde{N}}$ is parallel to $\bn$, which is perpendicular to $\bv_\perp$ (note that this does not hold at higher order in the velocity). On the contrary the effect is maximum for a binary which is edge-on, and with $\bv_\perp$ orthogonal to the plane of the binary, as illustrated in Fig.~\ref{fig:iota}. The fact that the amplitude of the effect depends on $\mathbf{\tilde{N}}$, i.e.\ on the orientation of the plane with respect to the observer, is directly linked to the fact that the amplitude of the polarisations scales with $\cos\tilde\iota$. For $\iota=0$, the change is quadratic in $\delta\iota$: $\cos(\tilde\iota)=\cos(0+\delta\iota)\simeq 1-\delta\iota^2/2$, whereas for $\iota=\pi/2$ the change is linear: $\cos(\tilde\iota)=\cos(\pi/2+\delta\iota)\simeq-\delta\iota$.

In Fig.~\ref{fig:iota} we show for illustration the case where the effect of aberration is maximum. In this configuration, if the source were not moving, we would receive only the $h_+$ polarisation, since only $h_+$ is emitted along $\bn$ ($\cos\iota=0$ meaning that $h_\times=0$). However, since the source is moving, we do not receive the GWs that have been emitted in direction $\bn$ but rather the GWs that have been emitted in direction $\tbn$ (and that we see coming from direction $\bn$). Along $\tbn$ both $h_+$ and $h_\times$ are produced and therefore we observe these two polarisations. From this we wrongly conclude that the plane of the binary is slightly inclined with respect to us, i.e.\ that the binary is not edge-on.

The second effect, the mixing of polarisations, simply means that the true polarisation of the source cannot be inferred, but that one measures instead a wrong polarisation
\begin{align}
\Psi=\tilde{\Psi}- v_1\frac{\cos\theta}{\sin\theta}\, .
\end{align}
Like for aberration, this means that the plane of the binary appears slightly turned (this time around $\bn$) with respect to the observer.

We see therefore that the source velocity biases the measurement of the angles describing the orientation of the binary system with respect to the observer. However, since these intrinsic parameters are unknown and randomly distributed over the population of sources, having a wrong measurement of them has no direct observational impact. In particular, the other parameters like the luminosity distance and the chirp mass are not affected by aberration and by the change in polarisation, since the source velocity is fully reabsorbed into the new direction $\bn$ and the new polarisation $\Psi$. This can be mathematically seen with the Fisher formalism. The measured strain $h$ depends on a set of parameters $\mathbf{\Theta}$. The Fisher matrix associated to these parameters is given by 
\begin{align}
\Gamma^{ij}=\left(\frac{\partial h(\mathbf{\Theta})}{\partial \Theta_i}\Big|\frac{\partial h(\mathbf{\Theta})}{\partial \Theta_j} \right)\, 
\end{align}
where the scalar product is defined as
\begin{equation}
    (a|b)=4 \mathcal{R}\mathrm{e} \left[ \int_{f_{\rm low}}^{\rm f_{\rm high}} \frac{\hat{a}(f)^* \hat{b}(f)}{S(f)} df\right],
\end{equation}
where $S(f)$ is the detector power spectral density (PSD), $f$ is the GW frequency, $*$ indicates the complex conjugate, $\hat{\cdot}$ the Fourier components, $f_{\rm low}$ is a low frequency cut-off given by the detector sensitivity and $f_{\rm high}$ an high frequency cut-off given by the sampling rate of data. The bias induced on the parameters $\mathbf{\Theta}$ by the source velocity is then given by 
\begin{align}
\Delta \Theta^i=(\Gamma^{-1})^{ij}\left(\frac{\partial h(\mathbf{\Theta})}{\partial \Theta_j}\Big| h(\mathbf{\Theta})-\tilde{h}(\mathbf{\Theta}) \right)\, ,
\end{align}
where $\tilde h$ is the strain that we would have in the absence of velocity. In our case, the difference between $h$ and $\tilde h$ can be fully reabsorbed into a different polarisation and different inclination angle. The observed strain~\eqref{eq:strain_hat} is found from the transformation rules~\eqref{eq:h_aberr} which are equivalent to \eqref{eq:RotationRab}, that is to a rotation of the source with $R$. Hence we can write that $\tilde h({\mathbf{\Theta}})=h(R^{-1}({\mathbf{\Theta}}))$, where $R^{-1}({\mathbf{\Theta}})$ are the parameters characterising a source with initial parameters $\mathbf{\Theta}$ and subsequently rotated with $R^{-1}$. That is, if $\mathbf{\Theta}$ defines a binary plane orthogonal to $N_i$, then $R^{-1}({\mathbf{\Theta}})$ defines a rotated binary plane orthogonal to ${R^{-1}_i}^j N_j$. Defining the parameters shifts by $\delta\mathbf{\Theta} = \mathbf{\Theta} - R^{-1}({\mathbf{\Theta}})$, the only non-zero components of $\delta\mathbf{\Theta}$ are $\delta\Psi$ and $\delta(\cos\iota)$ since they characterise the orientation of the binary plane. Taylor expanding around $\mathbf{\Theta}$ we then obtain
\begin{align}
\tilde h(\mathbf{\Theta})\simeq h(\mathbf{\Theta})-\frac{\partial h(\mathbf{\Theta})}{\partial \Theta^k}\delta\Theta^k\, ,
\end{align}
leading to
\begin{align}
\Delta \Theta^i&=(\Gamma^{-1})^{ij}\left(\frac{\partial h(\mathbf{\Theta})}{\partial \Theta_j}\Big| \frac{\partial h(\mathbf{\Theta})}{\partial \Theta_k} \right)\delta\Theta^k\\
&=(\Gamma^{-1})^{ij} \Gamma_{jk}\delta\Theta^k=\delta\Theta^i\nonumber\, .
\end{align}
Hence we see that the only parameters that are biased by the transverse velocity of the source are the polarisation and the inclination angle. In particular, the source transverse velocity has no impact on the luminosity distance and the chirp mass.

Let us conclude this section by noting that while the difference between $\bn$ and $\tbn$ depends on the relative velocity between the source and the observer and is therefore the same if the source moves with velocity $\bv_\perp$ with respect to the observer or if the observer moves with velocity $-\bv_\perp$ with respect to the source, the observational consequences are different in these two cases. In the case of a moving source, the incoming direction of the GW in the observer frame is not affected by the motion. As a consequence $\bn$ denotes the true direction of the source, and $\tbn$ is the direction of emission in the source frame, as depicted in Fig.~\ref{fig:iota}. The source velocities have therefore no impact on the observed position of sources in the sky. The velocity only affects the part of the source that the observer sees. On the other hand, in the case of a moving observer, the emitted direction of the GW in the source frame is not affected by the motion. Consequently $\tbn$ denotes the true direction of the source, and $\bn$ the apparent direction, seen by the moving observer. The observed positions of sources in the sky are therefore affected by the observer velocity. More precisely, the observer velocity with respect to a frame were sources are on average at rest generates a dipole in the source distribution, as computed for example in~\citet{Mastrogiovanni:2022nya} for GW events or in~\citet{Domenech:2022mvt} and \citet{Dalang:2022gfv} for galaxy counts. In the CMB, the observer velocity not only leaves a dipole which has been observed \citep{Kogut:1993ag,Lineweaver:1996xa}, but also distinctive off-diagonal correlations of both intensity and polarization which allow to constrain independently its magnitude \citep{2011PhRvL.106s1301K} and direction \citep{Amendola:2010ty,Planck:2013kqc,Mukherjee:2013zbi,Saha:2021bay}.

\section{Conclusions}\label{sec:conclusion}

In this paper, we showed that the peculiar motion of a gravitational wave source with respect to the observer rest frame, induces a distortion in the observed waveform. In particular the presence of a (non-zero) component of the peculiar velocity transverse to the line-of-sight gives rise to apparent vector polarisations in the observer frame. These are pure projection effects, proportional to the plus and cross polarisations in the source frame. They share therefore the same time dependence as the spin-2 modes and do not correspond to new degrees of freedom. We have shown that this implies that the spin-1 modes can always be rewritten as spin-2 modes coming from an aberrated direction, and with a slightly different polarisation. 

One could however wonder if by actively searching for vector modes, i.e. by including spin-1
antenna patterns in the modeling of the strain, one could
measure the amplitude of these new modes, as well as the
true (non aberrated) source location. Comparing this with the aberrated direction obtained from the time-delay, one could then measure
the transverse velocity. We showed that unfortunately, this is not feasible since, without knowing the peculiar velocity, there is no unique/preferred way of splitting the signal into spin-2 modes and spin-1 modes. The only meaningful solution is therefore the one with no spin-1 mode. This is indeed the only solution for which the direction inferred from the waveform and the direction inferred from time delay are the same. 

A direct consequence of the aberration of GW sources is that the parameters encoding the orientation of the binary system with respect to the observer are biased. For example, a binary that is edge-on, for which we should only detect a $h_+$ polarisation, will appear slightly inclined since we will receive both $h_+$ and $h_\times$ polarisations. The inclination angle and the polarisation angle that we measure are therefore not the true ones. Since these angles are unknown and are independent of other parameters, like the chirp mass or the luminosity distance, this bias has no direct impact on astrophysical or cosmological constraints inferred from GW measurements  such as \citet{2021JCAP...08..026F,2022arXiv220311680L,O3-cosmo,2022PhRvD.105f4030M,2022arXiv220309237I}. However, it might impact studies aiming at constraining the inclination distribution of binaries \citep{2022arXiv220400968V}.

We stress that the same effect is present in the case of an astrophysical source emitting spin-1 waves: if we look at the electric field emitted by such a source we find that the direction of propagation of the spin-1 wave is aberrated and that the only effect on the source parameters is an apparent rotation (i.e. the intrinsic angles defining the source orientation are biased).  
For example, for gamma-ray burst sources, if one defines an angle $\iota$ between the line of sight and the normal to the rotation plane, the effect of a transverse velocity is given by a change in the source orientation due to aberration given by Eq.\,(\ref{Effect:iota}), and a mixing of the two spin-1 polarisations of the emitted electromagnetic radiation.

We observe that our findings significantly differ from the conclusions of \citet{Torres-Orjuela:2018ejx}. The authors of this reference compute distortions in the antenna pattern function of a GW detector, induced by a peculiar motion of the observer frame with respect to the frame of emission. Such peculiar motion could be far from negligible for a binary system orbiting a supermassive black hole. They find that a velocity component orthogonal to the line of sight gives a non-monotonic modification of the detected amplitude of the wave. 
The authors argue that an additional rotation of the GW polarization in its plane,
which is not taken into account by aberration, is responsible for this effect. However, as we have explicitly shown here, and in agreement with appendix C of \citet{2016PhRvD..93h4031B}, the effect of a peculiar velocity can always be recast as a direction dependent rotation. 
This has a profound impact on the resulting signal. Indeed, contrary to what is concluded in~\citet{Torres-Orjuela:2018ejx}, we find that the impact of transverse velocities on GWs is completely analogous to the one on electromagnetic signals, i.e.\ it can be fully explained in terms of Lorentz transformations without the need of invoking additional corrections. We also find that transverse velocities do not produce spurious non-GR-like signals nor modifications in the source luminosity distance, unlike what is claimed in that reference 

In \citet{Torres-Orjuela:2020dhw} the authors detail the claim made in \citet{Torres-Orjuela:2018ejx} that when considering aberration and polarization rotation, the GW signal emitted changes in a way which cannot be reabsorbed into a redefinition of the source's intrinsic orientation. They deduce that this allows the detection of the constant motion of a source. This is in direct contradiction with what we have shown explicitly in this article. 
Their proof is based on the determination of the transformation matrix which relates multipoles in the source frame to multipoles in the observer frame. Their Eq.~(37) can be seen as such a transformation when using a multi-index notation $L = (\ell,m)$, as it is of the form  $\sum_L {\cal Y}^L_K H^L = {H'}^K$, with ${\cal Y}^L_K$ a matrix, similar to the $D_{\ell m,\ell' m'}$ of \citet{2008PhRvD..78d4024G} relating the Weyl scalar multipoles. When they reach their Eq.~(38), they claim that it is in contradiction with their initial assumption (33) that aberration is a remapping of the directions of the gravitational fields emitted. They therefore conclude that the effect of a boost cannot be simply described as a remapping of the source orientation, in contrast with what happens in electromagnetism. However, this conclusion is based on a series of mistakes that lead to Eq.~(39).  First, Eq.~(38) just determines the components of the transformation matrix, i.e. ${\cal Y}^{\ell,m+s}_{k,m} - \delta_k^\ell \delta^s_0 \propto C_s(\ell,k,m)$, where $C_s(\ell,k,m)$ are some given coefficients for $s=-1,0,1$. There is no contradiction when realizing that the r.h.s of Eq.~(38) is also of the form $\sum_L {\cal Y}^L_K H^L$, hence the extraction of coefficients is immediate when choosing sources such that $H^{\ell,m} = \delta^\ell_L \delta^m_M$.\footnote{To illustrate the contradiction, they consider the special case of a source with only $\ell=2$ and $m=2$ and obtain Eq.~(39). However when replacing by such particular source one must use $H^{\ell, m} = \delta^\ell_2 \delta^m_2$, hence in Eq.~(39) the first $1$ should be replaced by $\delta^2_k \delta^2_n$, and it lacks a factor $\delta^m_2$ in front of $C_0$, a factor $\delta^{m+1}_2$ in front of $C_+$ and a factor $\delta^{m-1}_2$ in front of $C_-$, that is Eq.~(39) allows to determine separately the individual components of the transformation matrix ${\cal Y}^{2,2}_{k,2-s} - \delta^2_k \delta_s^0\propto C_s(2,k,2-s)$ for $s=-1,0,1$.} In addition, the expressions of these coupling coefficients $C_s$ are also not correct.\footnote{It can be easily checked on a special case, using the notation and equation numbers of \citet{Torres-Orjuela:2020dhw}. Let us consider a source velocity along the azimuthal direction, which from Eqs.~(54) and (55) implies $C_\pm(\ell,k,m) = 0$ since $v_\pm =0$. If in addition the source has an azimuthal symmetry ($m=0$), then $C_0(\ell,k,0) = 0$, from the prefactor $m$ of expression (53). One would then deduce that no extra mode is generated in the observer frame. However in the simple case of a source with only $\ell=2$ and $m=0$ the observer must see a mode with $\ell=3$ and $m=0$ as detailed in section III.C.2 of \citet{2008PhRvD..78d4024G}.} This is related to the fact that the authors have exchanged $\partial_\phi$ and $\partial_\theta$ when obtaining their Eq.~(27) from their Eq.~(25). Furthermore, when using the angular momentum raising and lowering operators of their Eq.~(6) one finds $2 i \partial_\theta = \text{e}^{- i \phi} J_+ - \text{e}^{i \phi} J_-$, hence 
\begin{align}
\partial_\theta\left(H^{\ell m} {}_s Y^{\ell,m}\right) &= \frac{H^{\ell m}}{2} \Big(\sqrt{(\ell-m)(\ell+1+m)} \,{}_s Y^{\ell, m+1}\text{e}^{-i \phi } \nonumber\\
&- \sqrt{(\ell+m)(\ell+1-m)}\, {}_s Y^{\ell, m-1}\text{e}^{i \phi }\Big)\,.
\end{align}
This expression is not equal to $(H^{\ell m}/2) A(\ell,m) {}_s Y^{\ell,m}$, which is the result used in \citet{Torres-Orjuela:2020dhw}, because ${}_s Y^{\ell, m\pm1}\text{e}^{\mp i \phi } \neq {}_s Y^{\ell, m}$. Also they found that the angle $\alpha$, which is usually referred to as the spin phase, appears only at second order in the boost velocity, hence it was removed from their Eq.~(25). However the spin phase is due to the non-parallel transport of the spherical basis, and appears at linear order in the velocity, as seen in Eq.~(B25) of ~\citet{2008PhRvD..78d4024G}, and corresponds to the angle $\delta \psi$ in the present article.\footnote{In Eq.~(B25) of \citet{2008PhRvD..78d4024G}, we read a spin phase which expressed in our notation is $\chi = -\cot(\theta/2) v_1$. However, this has been obtained in application of Eq.~(3.12) of \citet{Newman:1966ub} which allows to obtain the spin phase for the Cartesian basis in the complex plane related by a stereographic projection. Since this Cartesian basis is related to the natural spherical basis by a rotation in the plane with an angle $\phi$ (see explanation after Eq.~(3.7) of the latter article) we rather find, using Eq.~\eqref{eq:ab1}, that the spin phase is $\delta \psi = \chi  + \tilde \phi - \phi = - \left(\cot(\theta/2) - \csc \theta \right) v_1 =- v_1 \cot \theta $, in agreement with Eq.~\eqref{mixing}.} These mistakes also explain why the authors of \citet{Torres-Orjuela:2020dhw} find in Eq.~(32) that all modes $\ell'$ are excited by a boost at linear order in the velocity from a mode $\ell$ in the source frame, whereas it is only the case for modes satisfying $\ell-1\leq \ell' \leq \ell+1$, see section III.C.2 of \citet{2008PhRvD..78d4024G}. All subsequent results, namely sections VI and VII of \citet{Torres-Orjuela:2020dhw}, are affected as they are based on the use of Eq.~(32). The exact same mistakes plague Eqs.~(7)-(10) of \citet{Torres-Orjuela:2020oxq} which are the basis of the subsequent analysis.

Note that the problem of finding how the multipoles of a signal are transformed by a boost, that is finding the ${\cal Y}^L_K$, has already been solved for spin-0 and spin-2 quantities in the context of CMB temperature and polarization in e.g. \citet{Challinor:2002zh}, \citet{Dai:2014swa}, \citet{Yasini:2017jqg} or appendix G of \citet{Cusin:2016kqx}. Both problems are related because the correlation of spin-1 fields with null momentum (the photons) generates a spin-2 structure, also with null momentum, which describes the linear polarization of a photon bath, see e.g. \citet{Pitrou:2019hqg}.
In any case, the crucial point is that since we are observing from a single position, we cannot see the change in the multipoles due to the source velocity. We can only sample the GW field in one direction, and as we have demonstrated in this paper, the boosted field in one direction is fully degenerate with the unboosted one in a different direction.

We conclude with a final remark: one might be tempted to assume that the large-scale correlations of the (cosmic-flow) velocity across the sky would induce correlations between the inclination angle of different sources. Measuring such a correlation, would then provide a direct way of measuring the transverse cosmological velocity. Unfortunately, the correlation of inclination angle turns out to be always vanishing: the change in inclination $\delta\iota=\tilde\iota-\iota$ does not depend directly on the transverse velocity ${\bf{v}}_{\perp}$ but on the projection of ${\bf{v}}_{\perp}$ on a random variable $\tilde{\mathbf{N}}$. This completely removes the correlation (see appendix \ref{Or_Corr} for details). We stress that even if we would correlate $\cos\iota$ with another quantity, e.g. galaxy number density, the correlation would also vanish. Aberration can therefore not be used to measure the transverse velocity of sources.

\section*{Acknowledgement}
We thank Nicola Tamanini for interesting discussions and exchanges, Ruth Durrer for valuable discussions and feedback during the early stage of this project, Clifford Will for comments on transformation properties in more general metric theories and Guillaume Faye for discussions on post-Newtonian expansions of waveforms. C.~B.~acknowledges support from the Swiss National Science Foundation and from the European Research Council (ERC) under the European Union’s Horizon 2020 research and innovation program (Grant agreement No.~863929; project title ``Testing the law of gravity with novel large-scale structure observables". The work of G.~Cu.~is funded by Swiss National Science Foundation (Ambizione grant \emph{Gravitational wave propagation in the clustered universe}) and by CNRS. 
For the purpose of open access, the authors have applied a Creative Commons Attribution (CC BY) licence to any Author Accepted Manuscript version arising from this submission.
\newpage 
\appendix

\section{Transformations of the metric tensor}\label{BoostH}

We now consider how metric perturbation in 4 dimensions transforms under boost of velocity $-\mathbf{v}$ (since the observer moves with velocity $-\mathbf{v}$ with respect to the source). The metric perturbation transforms as 
\begin{align}
h_{\mu\nu}&=\Lambda_{\mu}^{\,\,\alpha}\Lambda_{\nu}^{\,\,\beta}\widetilde{h}_{\alpha\beta}\,,\nn\\
h^{\mu\nu}&=\Lambda^{\mu}_{\,\,\alpha}\Lambda^{\nu}_{\,\,\beta}\widetilde{h}^{\alpha\beta}\,,
\end{align}
where 
\begin{align}
&{\Lambda^0}_0 = \gamma \,,\quad {\Lambda^0}_i = {\Lambda^i}_0 =- 
\gamma v_i\,,\nn\\
&{\Lambda^i}_j = \delta^i_j +
\frac{\gamma^2}{1+\gamma}v^i v_j\,, \nn\\
&{\Lambda_0}^0 = \gamma \,,\quad {\Lambda_0}^i = {\Lambda_i}^0 = 
\gamma v^i\,,\nn\\
&{\Lambda_i}^j = \delta_i^j +
\frac{\gamma^2}{1+\gamma}v_i v^j\,, 
\end{align}
with $\gamma^{-2}=1-v_i v^i$ and $\beta^2 \equiv v_i v^i$. 

Let us start with upper indices
\begin{align}
h^{00}&=\Lambda^{0}_{\,\,\alpha}\Lambda^{0}_{\,\,\beta}\widetilde{h}^{\alpha\beta}\,,\nn\\
h^{0i}&=\Lambda^{0}_{\,\,\alpha}\Lambda^{i}_{\,\,\beta}\widetilde{h}^{\alpha\beta}\,,\nn\\
h^{ij}&=\Lambda^{j}_{\,\,\alpha}\Lambda^{j}_{\,\,\beta}\widetilde{h}^{\alpha\beta}\,.
\end{align}
Now we assume that in the non-tilde frame  (frame comoving with the source) we are in TT gauge, implying $\widetilde{h}^{00}=\widetilde{h}^{0i}=0$. 
\begin{align}
h^{00}&=\Lambda^{0}_{\,\,m}\Lambda^{0}_{\,\,n}\widetilde{h}^{mn}=\gamma^2 v_m v_n \widetilde{h}^{mn} \,,\nn\\
h^{0i}&=\Lambda^{0}_{\,\,m}\Lambda^{i}_{\,\,n}\widetilde{h}^{mn}=-\gamma v_m \left(\delta^i_n +
\frac{\gamma^2}{1+\gamma}v^i v_n\right)\widetilde{h}^{mn}\,,\nn\\
h^{ij}&=\Lambda^{i}_{\,\,m}\Lambda^{j}_{\,\,n}\widetilde{h}^{mn}\nn\\
&=\left(\delta^i_m +
\frac{\gamma^2}{1+\gamma}v^i v_m\right)\left(\delta^j_n +
\frac{\gamma^2}{1+\gamma}v^j v_n\right)\widetilde{h}^{mn}\,.
\end{align}
At linear order in the velocity 
\begin{align}
h^{00}&=0 \,,\nn\\
h^{0i}&=-v_m \widetilde{h}^{mi}\,,\nn\\
h^{ij}&=\widetilde{h}^{ij}\,.
\end{align}
In flat space it follows that 
\begin{align}
h_{00}&=0 \,,\nn\\
h_{0i}&= v_m \widetilde{h}_{mi}\,,\nn\\
h_{ij}&=\widetilde{h}_{ij}\,.
\label{eq:h_ab_boosted}
\end{align}
The wave in the observer frame (without a tilde) is not in the TT gauge anymore. However, it is possible to fix the TT gauge with respect to the observer by performing the set of transformations detailed in the next appendix.

\section{Gauge transformation to purely spatial perturbation}\label{Gauge}

Suppose we have a general plane gravitational wave of the form
\begin{equation}
    h_{ab} = H_{ab} f( k^a x_a) \quad\mbox{with }\quad k^a k_a = 0.\nn
\end{equation}
Without loss of generality we choose spatial axes such that the gravitational wave is propagating in the $\hat{z}$ direction, so that $\hat{k}_a = (1, 0, 0, 1)$. We now consider a gauge transformation of the form 
\begin{align}
    \xi^a &= \Xi^a F( k^a x_a),
\end{align}
where $F(u)$ is the integral of $f(u)$, i.e., ${\rm dF}/{\rm d}u = f(u)$. A gauge transformation of this form leads to a transformation of the metric perturbation
\begin{align}
    h_{ab}^{\rm new} &= h_{ab}^{\rm old} - \partial_a \xi_b - \partial_b \xi_a \nonumber\\
    &=H_{ab}^{\rm new} f( k^a x_a), 
\end{align}
in which
\begin{align}
    H_{ab}^{\rm new} &= H_{ab} - \left(\begin{array}{cccc}2\Xi_0&\Xi_x&\Xi_y&\Xi_z+\Xi_0\\\Xi_x&0&0&\Xi_x\\\Xi_y&0&0&\Xi_y\\
    \Xi_z+\Xi_0&\Xi_x&\Xi_y&2\Xi_z \end{array}\right).
\end{align}
Making the choice
\begin{align}
    \Xi_a = \left(\frac{1}{2} H_{00}, H_{0x}, H_{0y}, -\frac{1}{2} H_{00} + H_{0z}\right)
\end{align}
reduces the metric perturbation to purely spatial form
\begin{align}
    H_{ab}^{\rm new} &=\left(\begin{array}{cccc}0&0&0&0\\
    0&H_{xx}&H_{xy}&-H_{0x}\\
    0&H_{xy}&H_{yy}&-H_{0y}\\
    0&-H_{0x}&-H_{0y}&H_{zz}+H_{00}-2H_{0z}
    \end{array}\right).
\end{align}
For the particular metric components given in Eq.~(\ref{eq:h_ab_boosted}), this gauge transformation gives
\begin{align}
    &H_{ab}^{\rm new}=\nn\\ &\left(\begin{array}{cccc}0&0&0&0\\
    0&\tilde{h}_{xx}&\tilde{h}_{xy}&-v_x\tilde{h}_{xx}-v_y\tilde{h}_{xy}\\
    0&\tilde{h}_{xy}&\tilde{h}_{yy}&-v_x\tilde{h}_{xy}+v_y\tilde{h}_{xx}\\
    0&-v_x\tilde{h}_{xx}-v_y\tilde{h}_{xy}&v_y\tilde{h}_{xx}-v_x\tilde{h}_{xy}&0
    \end{array}\right) \label{eq:hTTobs}
\end{align}
in which we have used $\tilde{h}_{yy} = -\tilde{h}_{xx}$ and $\tilde{h}_{00}=\tilde{h}_{0z}=\tilde{h}_{zz}=0$. We see that this gauge transformation makes the purely spatial part of the metric equal to the electric components of the Riemann tensor, justifying the assumptions made in Section~\ref{sec:pattern}.

\section{Relation between proper times}
\label{app:dtau}

To relate the proper time in the source frame $d\tilde{\tau}$ to the proper time in the observer frame $d\tau$, we proceed in the following way. We first relate the 4-momentum of the GW in the source and observer frame to the phase $\Phi$~\footnote{Since $\Phi$ is a scalar, it is invariant under a boost: $\tilde{\Phi}=\Phi$. }
\begin{align}
\tilde{k}_{\mu}&=-\frac{\partial}{\partial \tilde{x}^\mu}\Phi\\
k_{\mu}&=-\frac{\partial}{\partial x^\mu}\Phi\, .
\end{align}
Since GWs propagate along null geodesics, the phase is conserved during propagation:
\begin{align}
\tilde{k}^\mu \tilde{k}_\mu=-\tilde{k}^\mu \frac{\partial}{\partial \tilde{x}^\mu}\Phi=0\, .
\end{align}
Let us now consider two GWs emitted subsequently: the first one at time $\tilde\tau$ with phase $\Phi(\tilde\tau)$ and the second one at time $\tilde\tau+d\tilde\tau$ with phase $\Phi(\tilde\tau+d\tilde\tau)$. The observer receives these GWs at time $\tau$ and $\tau+d\tau$ respectively and since the phase is conserved we have
\begin{align}
\Phi(\tilde\tau+d\tilde\tau)-\Phi(\tilde\tau)=\Phi(\tau+d\tau)-\Phi(\tau)\, .
\end{align}
Using that
\begin{align}
\Phi(\tilde\tau+d\tilde\tau)-\Phi(\tilde\tau)= d\tilde\tau\frac{d\Phi}{d\tilde\tau}
=d\tilde\tau\tilde{u}^\alpha\frac{\partial\Phi}{\partial\tilde{x}^\alpha}
=d\tilde\tau\tilde{u}^\alpha \tilde{k}_\alpha=-\tilde{E}d\tilde\tau\, ,
\end{align}
and similarly at the observer, we find
\begin{align}
\tilde{E}d\tilde\tau=Ed\tau\, .
\end{align}

\section{Vector and tensor mode splitting} \label{vec}
One could  wonder if by actively searching for vector modes, i.e.\ by including spin-1 antenna patterns in the modelling of the signal, one could measure the amplitude of these new modes, as well as the true direction $\tbn$. Comparing this with the aberrated direction obtained from the time-delay, one could then measure the transverse velocity $\mathbf{v}_\perp$. We show here that this turns out to be impossible, since there is no unique way of splitting the signal into spin-2 modes and spin-1 modes.

We start from the dimensionless driving force matrix computed in Eq.~\eqref{eq:hijGen} and we split the transverse velocity of the source as
\begin{align}
\bv_\perp=(\bv_\perp-\bw_\perp)+\bw_\perp\, ,
\end{align}
where $\bw_\perp$ is an arbitrary transverse velocity with amplitude $w_\perp\ll 1$ such that we can work at linear order in the velocities. The aberrated direction related to the transverse velocity $\bw_\perp$ is given by
\begin{align}
\mathbf{s}=\tbn+w_1\tilde{\mathbf{e}}_1+w_2\tilde{\mathbf{e}}_2\, .
\end{align}
Following Eqs.~\eqref{eq:eall} we define the two natural polarisation axes associated to $\mathbf{s}$:
\begin{subequations}
\begin{align}
\hat{\mathbf{f}}_1&=\tilde{\mathbf{e}}_1-w_1\tbn-w_1\frac{\cos\tilde{\theta}}{\sin\tilde{\theta}}\tilde{\mathbf{e}}_2\, ,\\
\hat{\mathbf{f}}_2&=\tilde{\mathbf{e}}_2-w_2\tbn+w_1\frac{\cos\tilde{\theta}}{\sin\tilde{\theta}}\tilde{\mathbf{e}}_1\, .
\end{align}
\end{subequations}
Inserting this into~\eqref{eq:hijGen} we find for the dimensionless driving force matrix
\begin{align}
P_{ij}&=\hat{H}_{+}\big(\hat{f}_{1i}\hat{f}_{1j}-\hat{f}_{2i}\hat{f}_{2j} \big) 
+\hat{H}_{\times}\big(\hat{f}_{1i}\hat{f}_{2j}+\hat{f}_{2i}\hat{f}_{1j} \big) \\
&+\hat{H}_1\big(s_{i}\hat{f}_{1j}+\hat{f}_{1i}s_{j} \big)
+\hat{H}_2\big(s_{i}\hat{f}_{2j}+\hat{f}_{2i}s_{j} \big)\, ,\nn
\end{align}
where
\begin{subequations}
\begin{align}
\hat{H}_+ &= \thh_+-2w_1\frac{\cos\tilde{\theta}}{\sin\tilde{\theta}}\thh_{\times} \label{eq:ha}\, ,\\
\hat{H}_{\times} &= \thh_{\times}+2w_1\frac{\cos\tilde{\theta}}{\sin\tilde{\theta}}\thh_{+}\, ,\\
\hat{H}_1 &= -(v_1-w_1) \thh_+ - (v_2-w_2) \thh_\times\, ,\\
\hat{H}_2 &= -(v_1-w_1) \thh_\times + (v_2-w_2) \thh_+ \label{eq:hb}\, .
\end{align}
\end{subequations}
From this we see that there is an infinite number of ways of splitting the signal into spin-2 and spin-1 modes with associated aberrated direction $\mathbf{s}$. There is no way to determine which splitting corresponds to the true velocity $\bv$ and therefore the only meaningful solution is the one with no spin-1 modes. This is indeed the only solution for which the direction inferred from the waveform and the direction inferred from time delay are the same.

\section{Zero correlation of orientation across the sky}\label{Or_Corr}

In this appendix we schematically prove that the 2-point correlation function of the source orientation across the sky is vanishing. 
We assume to have two pixels across the sky, each one containing a set of GW binary systems, with random orientations $\tilde N_1^i$ in the first pixel and $\tilde N_2^i$ in the second one. 

When we correlate two different pixels in the sky we get
\begin{align}
\langle \cos\tilde{\iota}_1 \cos\tilde{\iota}_2\rangle=&-\langle \cos\iota_1 \tilde{\bf{N}}_2 \cdot {\bf{v}}_{\perp}^2\rangle-\langle \cos\iota_2 \tilde{\bf{N}}_2 \cdot {\bf{v}}_{\perp}^1\rangle\nn\\
&+\langle \cos\iota_1 \cos\iota_2\rangle+\langle \tilde{\bf{N}}_1 \cdot {\bf{v}}_{\perp}^1 \tilde{\bf{N}}_2\cdot{\bf{v}}_{\perp}^2\rangle\,,
\end{align}
where the mean has to be interpreted as an ensemble average when acting on stochastic velocities and as a geometric mean over a bunch of sources when acting on geometric quantities. This can be rewritten as
\begin{align}\label{corr_o}
\langle \cos\tilde{\iota}_1 \cos\tilde{\iota}_2\rangle =&-\langle \cos\iota_1 \rangle \langle \tilde{\bf{N}}_2 \cdot {\bf{v}}_{\perp}^2\rangle-\langle \cos\iota_2 \rangle \langle \tilde{\bf{N}}_2 \cdot {\bf{v}}_{\perp}^1\rangle\nn\\
&+\langle \cos\iota_1 \rangle \langle\cos\iota_2\rangle+\langle \tilde{N}^i_1 \tilde{N}^j_2\rangle  \langle v_{\perp_i}^1 v_{\perp_j}^2\rangle\,.
\end{align}
It is apparent that the first three terms on the right hand side vanish. However the last vanishes as well due to
\be
\langle \tilde{N}^i_1 \tilde{N}^j_2\rangle=\langle \tilde{N}^i_1\rangle \langle\tilde{N}^j_2\rangle=0\,,
\ee
which states that the orientation in two different pixels is not correlated, and the orientations inside each pixel are randomly distributed.

Notice that if one takes the limit $1\rightarrow2$ in (\ref{corr_o}), it appears that cosmological velocities give a  modification in the variance of the velocity field. This is due to the fact that when computing the aberration angle, we kept only the first order term in the velocity. However, we need to make sure that unit vectors have unit norm, as an incorrect normalization brings biases when estimating the variance of $\cos \iota$. Explicitly, for the aberrated direction, one has to consider 
\begin{equation}
\tbn = \frac{\bn + \bv_\perp}{\sqrt{1+ v_\perp^2}}\,.
\end{equation}
Now we use that the average of a direction vector is such that $\langle \tilde N_i \tilde N_j \rangle = (1/3) \delta_{ij}$ hence $\langle (\cos \iota)^2 \rangle = (\bn \cdot \bn)/3 = 1/3$.
Then we compute
\begin{equation}
\langle (\cos \tilde \iota)^2 \rangle = \frac{\langle (\cos \iota)^2 \rangle + \langle (\tilde N \cdot \bv_\perp)^2\rangle }{1+ v_\perp^2} = \frac{\frac{1}{3}+ \frac{v_\perp^2}{3}}{1+ v_\perp^2} = \frac{1}{3}\,.
\end{equation}
Hence $\langle (\cos \tilde \iota)^2 \rangle = \langle (\cos \iota)^2 \rangle$, showing that the variance of the orientation is also not affected by a global velocity flow.

\bibliography{refs}
\bibliographystyle{mnras}
\label{lastpage}

\end{document}